\documentclass[%
 aip,
 amsmath,amssymb,
 reprint,%
]{revtex4-1}
\usepackage{graphicx}
\usepackage{subcaption}
\usepackage{amsmath}
\usepackage{xcolor}
\usepackage{soul}

\usepackage{mathptmx}
\usepackage{etoolbox}
\usepackage[utf8]{inputenc}
\usepackage[T1]{fontenc}
\usepackage{bm}

\draft

\makeatletter
\def\@email#1#2{%
 \endgroup
 \patchcmd{\titleblock@produce}
  {\frontmatter@RRAPformat}
  {\frontmatter@RRAPformat{\produce@RRAP{*#1\href{mailto:#2}{#2}}}\frontmatter@RRAPformat}
  {}{}
}%
\makeatother

\begin{document}

\title[]{The model of photoswitching in lead-halide perovskite  microcrystals}
\author{Eduard A. Podshivaylov}
\author{Pavel A. Frantsuzov }
\email{frantsuzov@kinetics.nsc.ru}
\affiliation{Voevodsky Institute of Chemical Kinetics and Combustion SB RAS, 630090, Novosibirsk, Russia}%

\date{\today}

\begin{abstract}
A theoretical model for the recently discovered effect of all-optical photoswitching in lead tribromide perovskite single microcrystals is proposed. The model takes into account the spatially distributed kinetics of the charge carrier recombination and the creation/destruction of trap states. It successfully reproduces the key features of the photoswitching phenomenon.
\end{abstract}

\maketitle

\section{Introduction}
Metal-halide perovskite semiconductors have received a great deal of attention in the past decade. Perovskites, which have the general chemical formula of ABX\textsubscript{3}, where the most common choices are methylammonium, cesium or formamidinium for A, lead or tin for B, and bromine, chlorine or iodine for X, can be formed into a large number of systems of various scales and dimensions - from quantum dots to granular films and whole crystals. Due to their unique photophysical properties, these systems open the way for a wide range of different applications, including quantum sensors \cite{ShellaiahChemoSensors2020}, phosphors \cite{WeiChemSocRev2019, LuoACSapMat2019}, photodetectors \cite{WangADVsc2021, TianSmall2017}, and photovoltaic elements \cite{JenaChemRew2019, ChenAccChemRes2021}.

Some of the most unusual properties of perovskites are the variation of their crystal structure over time and their great tolerance to defects. These properties give rise to such effects as luminescence enhancement \cite{ChenSolRRL2017, SamuelAdvEnMar2020}, aging \cite{MinussiSusEnFuels2022}, and "self-healing" \cite{Ghosh2020, Cahen2021, Li2021} of the individual crystals, halide ion photoinduced separation in mixed perovskites \cite{BischakNL2017, Draguta2017, Barker2017}, luminescence quenching and recovery under the effect of external mechanical pressure \cite{GalleAdvSc2023}, single microcrystal \cite{Seth2021, Pathoor2018} and quantum dot \cite{Han2020, Li2018, SethJPC2016} luminescence blinking, ion migration \cite{LiuAdvFunMat2021, Azpiroz2015,Yuan2016,Panzer2017,Stecker2019}, and others.

\begin{figure}[h]
\includegraphics[width=1\linewidth]{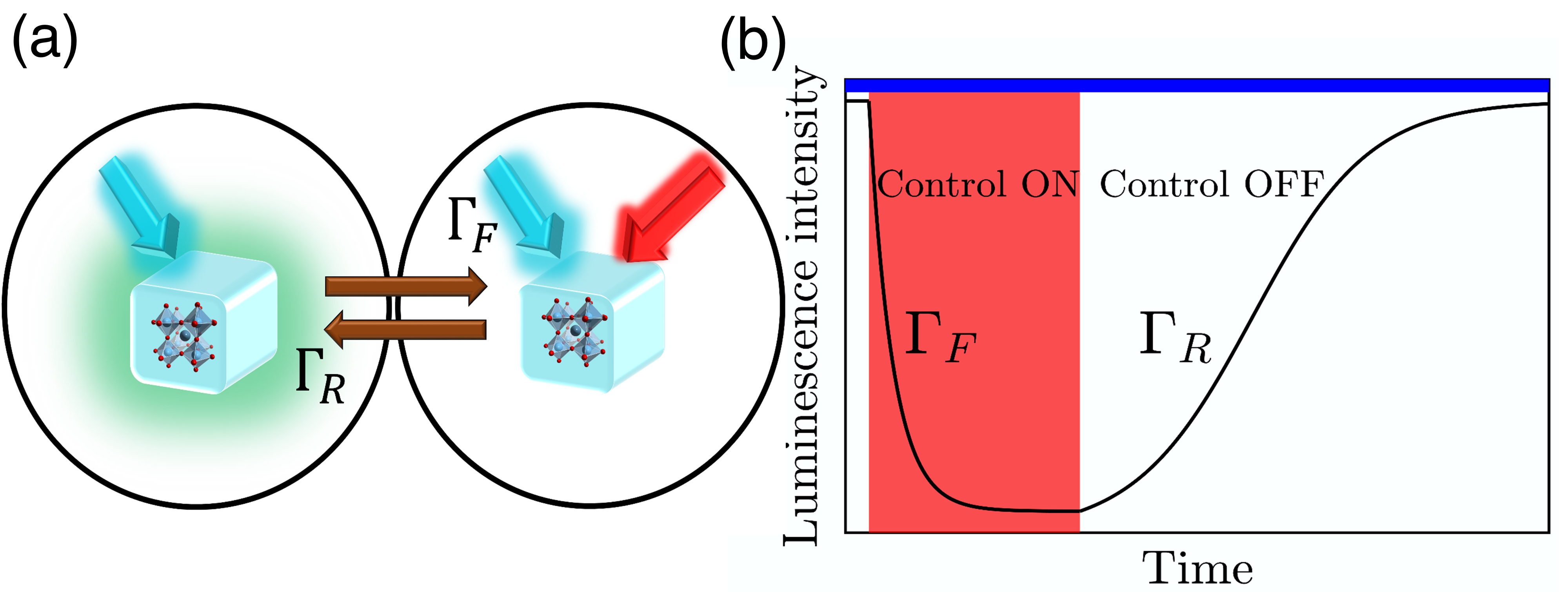}
\centering
\caption{(a) General picture of a single perovskite microcrystal photoswitching under control light irradiation. (b) Characteristic time dependence of the photoinduced luminescence switching (black line), the periods of excitation and control light irradiation are marked with blue and red colors correspondingly.}
\label{GenPic}
\end{figure}

Recently, the photoinduced luminescence quenching phenomenon was discovered for single MAPbBr$_3$,  FAPbBr$_3$, and CsPbBr$_3$ perovskite microcrystals with lead excess \cite{TianAdvMat2023}. Luminescence is caused by an excitation light, while the quenching is caused by a control light which has a photon energy less than the crystal band gap energy. However, the luminescence is restored after the control light is turned off, and therefore one can switch the crystal luminescence on and off. The general picture of this phenomenon is presented in Fig. \ref{GenPic}.

The phenomenon was experimentally found to exhibit the following important properties:

1. The shape of the luminescence quenching curve under the control light is well approximated by the exponential law:
\begin{equation}
PL(t) = (PL_{ON}-PL_{OFF}) \exp[-\Gamma_F (t-t_{ON})] +PL_{OFF},
\label{ONdep}
\end{equation}
where $t_{ON}$ is the control light switching on time, $\Gamma_F$ is the falling rate, $PL_{ON}$ and $PL_{OFF}$ are the photoluminescence intensities in the ON and OFF states respectively.

2. The luminescence intensity dependence after switching off the control light is well approximated by the logistic curve:
\begin{equation}
PL(t) = \frac{ PL_{ON}PL_{OFF}\exp[\Gamma_R (t-t_{OFF} )]}{PL_{ON}+PL_{OFF} (\exp[\Gamma_R (t-t_{OFF})]-1)},
\label{OFFdep}
\end{equation}
where $\Gamma_R$ is the rising rate of photoluminescence intensity.

3. The falling rate $\Gamma_F$ and the switching efficiency
\begin{equation}
SE= \frac{PL_{ON}-PL_{OFF}}{PL_{ON}}
\end{equation}
depend on the excitation and control light intensities.The switching efficiency has an S-shaped curve with saturation. Both the SE and the falling rate increase with an increase in the control light intensity and decrease with an increase in the excitation light intensity. The control light also affects the crystal in the absence of excitation light.

4. The rising rate of the luminescence $\Gamma_R$ linearly depends on the intensity of the excitation light, but also depends on the SE, \textit{i.e.} the rising rate decreases with the SE increase. This means that the influence of the control light does not disappear immediately when it is turned off, but with a delay that depends on its intensity.

5. The average luminescence lifetime decreases when the control light is turned on. The maximal lifetime is in order of 100-200 ns when the crystal is not affected by the control light and can decrease up to 10 ns under the control light illumination.

6. The falling rate has a clear Arrhenius temperature dependence with a characteristic energy of about 70 mEv. The rising rate, however, is practically independent of temperature.

7. The rising rate strongly depends on the wavelength of the excitation light and correlates with the excitation light absorption of the crystal.

In the original article of Wan \textit{et al.} \cite{TianAdvMat2023} a model that described most of the key properties was presented. The model was based on the idea of strong non-radiative recombination centers that are directly created and annihilated under illumination by a control and excitation light, respectively. Nonradiative centers are associated with defects of the perovskite crystal structure. The proposed model corresponds to the following kinetic scheme:
\begin{eqnarray}
    &P + h\nu_c \rightarrow A,\nonumber\\
    &A + h\nu_e \rightarrow P.\nonumber
\end{eqnarray}

Here $P$ stands for the density of the passive centers, which can be turned into active centers and $A$ stands for the density of the active centers. In the original work, the authors also observed an additional effect of the luminescence enhancement over time, and therefore they added another type of nonradiative centers to describe this process. This process is slow compared to photoswitching and is outside the focus of this article. However, by taking this effect into account, the original work's authors were unable to propose an analytical solution to their model.

Using the fact that the sum density is a constant $A(t) + P(t) = A_M$ the model can be described by the following equation:

\begin{equation}
    d_t A(t) = \sigma_c J_c [A_M - A(t)] - \sigma_E J_e A(t),
    \label{OrigModel}
\end{equation}

where $\sigma_c$ is the effective absorption cross-section of the defect creation process under control light illumination and $\sigma_E$ is the effective absorption cross-section of the defect annihilation process under excitation light illumination. $J_c$ and $J_e$ are the control light and excitation light photon fluence rates, respectively.
 $$J_c \equiv \frac{I_c}{\hbar\omega_c}, \: J_e \equiv \frac{I_e}{\hbar\omega_e},$$
where $I_c$ and $I_e$ are the control light and excitation light intensities, respectively, and $\hbar\omega_c$ and $\hbar\omega_c$ are the control light and excitation light photon energies, respectively.

The stationary density of active defects with no control light illumination is $A=0$. Once the control light is switched on the solution of Eq. (\ref{OrigModel}) is:

\begin{equation}
    A(t) = A_0
    \left\{1-\exp\left[-\left(\sigma_c J_c +\sigma_E J_e \right)(t-t_{ON})\right]\right\},
    \label{OrigON}
\end{equation}
where
\begin{equation}
    A_0=A_M \frac{\sigma_c J_c}{\sigma_c J_c +\sigma_E J_e}
\end{equation}
is the long-term limit of the active defects density, \textit{i.e.} the density that corresponds to the minima of the PL curve. After the control light has been switched off, the defect density decreases:

\begin{equation}
    A(t) = A_0 \exp\left[-\sigma_E J_e (t-t_{OFF})\right].
    \label{OrigOFF}
\end{equation}

To calculate the resulting luminescence intensity the authors suggested the Stern-Volmer law of luminescence quenching:

\begin{equation}
    PL (t) = \frac{PL_{ON}}{1+Q A(t)},
    \label{Stern}
\end{equation}
where $Q$ is the Stern-Volmer constant.

Combining the equations (\ref{OrigON}-\ref{OrigOFF}) and the Stern-Volmer law (\ref{Stern}), one can reproduce the characteristic photoswitching time dependence (\ref{ONdep}-\ref{OFFdep}), which corresponds to Properties 1 and 2. It is also possible to reproduce Properties 3 and 7. Property 6 can be explained if one assumes $\sigma_c$ to have the Arrhenius-modified form of $\sigma_c=\sigma_0 \exp(-\Delta E/k_B T)$ with some barrier $\Delta E$ and pre-exponential factor $\sigma_0$. Nevertheless, the model has a number of significant drawbacks.

{\bf The first one} is neglect of the strong extinction of the excitation. Most of the experiments in the original work were performed under 450 nm wavelength excitation. The extinction coefficient $ \kappa $ of MAPbBr\textsubscript{3} for this wavelength was experimentally found by Ishteev \textit{et al.} \cite{IshteevJMatChemC2022} and roughly equals to 0.197. The corresponding extinction length is then defined as $l_e=\lambda_e/4 \pi \kappa \approx 200$ nm, which is much less than the characteristic crystal length $L_{cr}$ of 10 $\mu$m. This means that the excitation light is absorbed in a very thin initial layer of the crystal, and absolutely all of the excitation light illuminating the crystal is absorbed. At the same time, the control light extinction length is at least one order of magnitude larger. Therefore, it becomes clear that the control light could be the cause of the defects. The excitation light, however, cannot directly affect them in any way, but is rather associated with their annihilation through the recombination of charge carriers.

{\bf The second drawback} stems from the first. If one must consider the charge carrier dynamics, it is then crucial to account for the carrier diffusion. The electron diffusion coefficient is for these systems is in order of  $D_n \approx  1$ cm\textsuperscript{2}/s \cite{Elbaz2017, Zhumekenov2016, Shi2015, Scajev2020, Oksenberg2021}. By combining the diffusion coefficient with the average carrier's lifetime it is possible to estimate the diffusion length. Assuming that the luminescence lifetime $\tau_L=100$ ns is equal to the electron lifetime, we can estimate the maximal diffusion length as $L_D = \sqrt{D_n \tau_L} \approx 1$ $\mu$m, which is consistent with the values obtained in the references above. The value of the diffusion length is less than the characteristic microcrystal's length of 5 to 10 $\mu$m, and thus the carrier diffusion must be explicitly taken into account. Note that the hole diffusion coefficients in MAPbBr$_3$, FAPbBr$_3$ and CsPbBr$_3$ are several times larger than the electron diffusion coefficients.

{\bf The third drawback} is the use of the Stern-Volmer law. The authors did not use a specific kinetic model for the charge carrier recombination, but applied the Stern-Volmer equation to estimate the luminescence quenching dependence on the number of active centers. The Stern-Volmer equation is used to determine the luminescence quenching of molecules in the presence of extinguishers, but cannot generally describe the luminescence intensity of a semiconductor crystal.

As such, it is necessary to build a model that would be free from the drawbacks of the previously proposed model and which could reproduce the photoswitching properties described above. Such a model should consist of two parts - the first should take into account the fast charge carrier recombination kinetics (on a timescale of hundreds of nanoseconds),  while the second should describe the slow dynamics of defect creation and annihilation on a timescale of milliseconds and seconds.

\section{Charge carrier recombination kinetics}
\subsection{Basic equations}
\begin{figure}[h]
\includegraphics[width=1\linewidth]{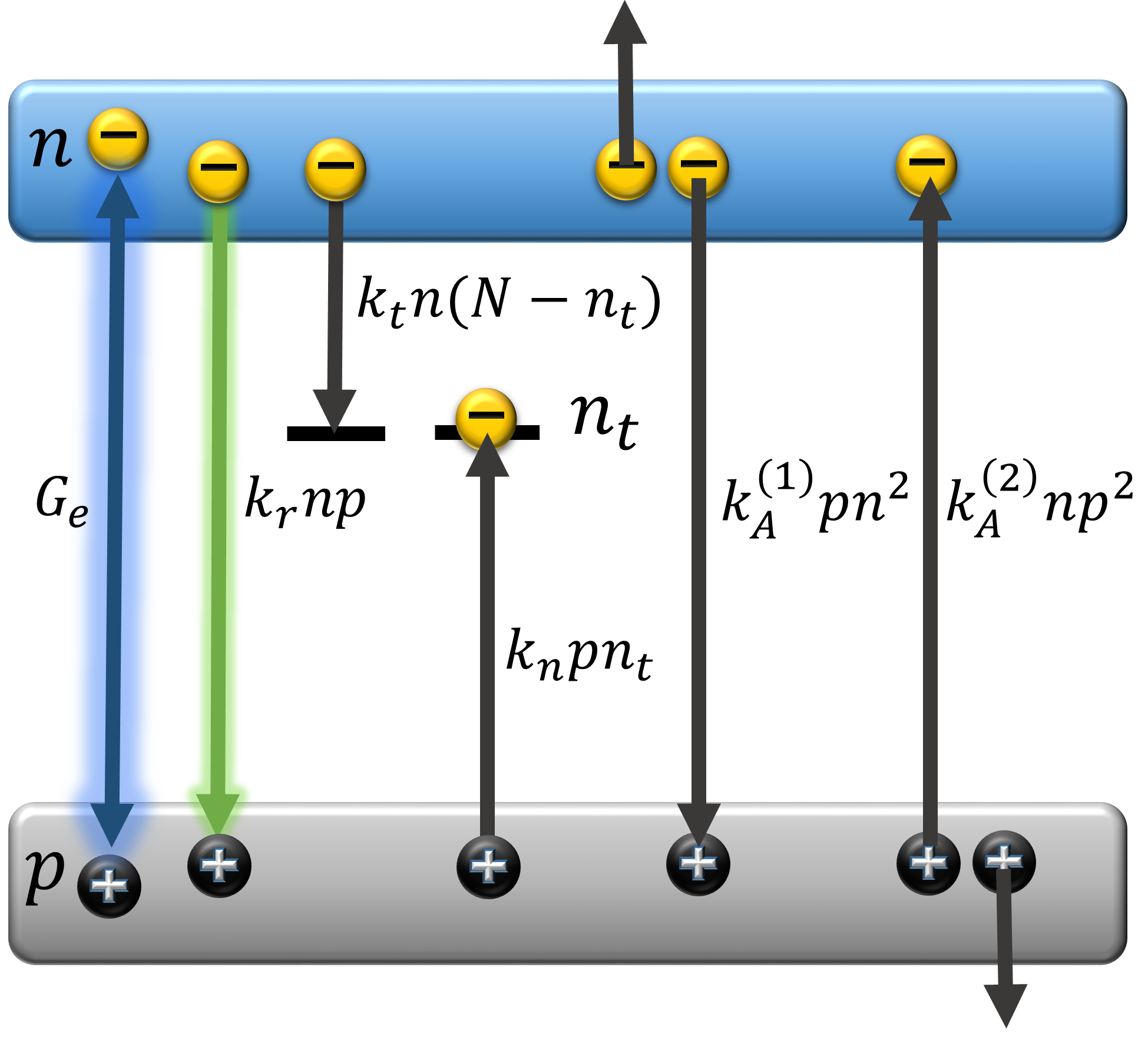}
\centering
\caption{ The schematic picture of the proposed model. Only two of the six Auger processes are shown.}
\label{GenKin}
\end{figure}

The semiconductor crystal luminescence is determined by the kinetics of photoexcited charge carriers. Carrier recombination kinetics in semiconductors are traditionally described by the Shokley-Reed-Hall (SRH) model \cite{Shockley1952,Hall1952}, where the nonradiative recombination of the charge carriers occurs via trapping into the state in the band gap. Generalizations of this model, including for a few types of trap states (see Ref. \onlinecite{Vietmeyer2011}, for example), have been proposed. At high excitation power Auger recombination processes have to be taken into account, as it was done in the ABC model \cite{Dai2010}. The modified Shokley-Reed-Hall model (SRH+) suggested recently \cite{Kiligaridis2021} was successfully used to describe the charge carrier kinetics of various perovskite films. Here we use a more generalized SRH+ model that takes into account the spatial distribution and diffusion of free carriers. The schematic picture of the general kinetic model is shown in Fig. \ref{GenKin}. It corresponds to the following set of equations:

\begin{equation}
\begin{cases}
        \partial_t  n = \nabla [D_n \nabla n +e \mu_n n \nabla \varphi]  \\
      \qquad + G_e(\vec{r},t) - k_t [N(\vec{r})-n_t] n -k_r n p - R_n\\
      \partial_t  n_t = k_t [N(\vec{r})-n_t] n - k_n n_t p - R_t \\
      \partial_t  p= \nabla [D_p \nabla p -e \mu_p p \nabla \varphi]  \\
      \qquad   +G_e(\vec{r},t) - k_n n_t p - k_r n p - R_p.
\end{cases}
\label{SRH+}
\end{equation}

Here $G_e(\vec{r},t)$ is the carrier generation rate, which is a function of time and space and depends on the pump time profile and the absorbing properties of the medium; $n(\vec{r},t),\: p(\vec{r},t),\: n_t(\vec{r},t)$ are the densities of the electrons, holes, and trapped electrons, respectively, $\varphi(\vec{r},t)$ is the electric field potential induced by the charge carriers, $N(\vec{r})$ is the trap density, $k_t$ is the rate constant of electron trapping, $k_n$ is the rate constant of the hole non-radiative recombination with a trapped electron, and $k_r$ is the radiative rate constant, $e$ is the elementary charge. $\mu_n$ and $\mu_p$ are the mobility, $D_n = \mu_n k_B T$ and $D_p = \mu_p k_B T$ are the diffusion coefficients of electrons and holes, respectively. Functions $R_n,\: R_t,\: R_p$ are the third-order multivariate polynomials that correspond to the different Auger processes occurring in the crystal. We assume the diffusion to be isotropic. In general, the diffusion tensor must be used in equation (\ref{SRH+}), as was done, for example, in Ref. \onlinecite{deQuilettes2022}.

The system (\ref{SRH+}) must be supplemented by the Poisson equation for the electric potential:

\begin{equation}
    \varepsilon \varepsilon_0 \Delta \varphi = -e [p(\vec{r},t)-n(\vec{r},t)-n_t(\vec{r},t)],
    \label{field_eq}
\end{equation}
where $\varepsilon$ is dielectric permittivity, $\varepsilon_0$ is vacuum permittivity.

The system of equations must be completed with initial and boundary conditions. The initial conditions correspond to the absence of charges before any illumination occurs:

\begin{equation}
n(\vec{r},0)=n_t(\vec{r},0)=p(\vec{r},0)=0.
\end{equation}

The boundary conditions correspond to the absence of holes' and electrons' flux through the crystal surface:

\begin{equation}
    -D_n  \left.\nabla_{e} n (\vec{r},t)  \right|_\text{ surf} = -D_p \left.\nabla_{e} p (\vec{r},t)  \right|_\text{ surf} = 0,
    \label{Bound}
\end{equation}
where $\nabla_e$ is the normal component of the gradient vector.

The boundary conditions for the electrical potential:

\begin{equation}
 \left.\nabla_e \varphi (\vec{r}, t)\right|_\text{ surf}  =  \left.\nabla_\tau \varphi (\vec{r},t)\right|_\text{ surf} = 0,
 \label{field_bound}
\end{equation}
where $\nabla_\tau$ is the tangential component of the gradient vector.

The equations of system (\ref{SRH+}) are not independent because of the law of charge conservation. As such, we will use the form:

\begin{equation}
    \int n(\vec{r},t) d^3r + \int n_t(\vec{r},t) d^3r = \int p(\vec{r},t) d^3r.
    \label{Charge}
\end{equation}

The measured interband luminescence emission power for such a model can be expressed by the following expression:

\begin{equation}
    PL(t) =  \hbar \bar{\omega} k_r \int n(\vec{r},t) p(\vec{r},t) F(\vec{r}) d^3r,
\end{equation}
where $\hbar \bar{\omega}$ is the average emitted photon energy,  $F(\vec{r})$ is the emission survival probability, which takes into account the reabsorption of light by the crystal.

The resulting system of equations (\ref{SRH+}-\ref{field_eq}) is difficult to solve in its complete form. However, due to the specific experimental conditions, we can apply a few approximations that will simplify the problem:

\textbf{a)} We will consider the excitation light pump to be continuous. As such, we cannot consider Property 5 in our model. Such an approximation, however, allows us to disregard the time derivative included in the system (\ref{SRH+}).

\textbf{b)} Most of the experiments in the original work were carried out using an intensity of 1 W/cm$^2$ for the excitation light pump. At such an intensity, the estimation of the maximal average electron density is $n_{\max}= G_e \tau_L \approx J_e \tau_L /L_{cr} = $ 10$^{15}$ cm$^{-3}$. As such,
It was shown by number of experimental works (see, for example, Refs. \onlinecite{Kiligaridis2021, Droseros2018}) that at such excitation intensity the Auger processes do not significant in recombination kinetics.
So we can neglect them in the system (\ref{SRH+}).

Thus, with all of these assumptions in place, the system (\ref{SRH+}-\ref{field_eq}) can be rewritten in the following form:
\begin{equation}
\begin{cases}
     \nabla [D_n \nabla n (\vec{r}) +e \mu_n n (\vec{r}) \nabla \varphi (\vec{r})]\\
    \qquad  =   k_t [N(\vec{r})-n_t] n +k_r np - G_e(\vec{r}) \\
     k_n n_t(\vec{r}) p(\vec{r}) = k_t [N(\vec{r})-n_t] n   \\
     \nabla [D_p \nabla p(\vec{r}) -e \mu_p p (\vec{r}) \nabla \varphi (\vec{r})]\\
    \qquad =   k_n n_t p +k_r np - G_e(\vec{r}) \\
     \varepsilon \varepsilon_0 \Delta \varphi (\vec{r})= -e [p(\vec{r})-n(\vec{r})-n_t(\vec{r})],
\end{cases}
\label{SRHsimple}
\end{equation}
with the boundary conditions (\ref{Bound}-\ref{field_bound}).

\subsection{Rapid diffusion limit}

In the rapid diffusion limit, where the diffusion length far exceeds the size of the crystal, the charge carriers are homogeneously distributed within the crystal. Mathematically, this means that their densities do not depend on $\vec r$. Uniform distribution of charges results in the absence of an electric field inside the crystal. In this limit, the system (\ref{SRHsimple}) has the following form:
\begin{equation}
\begin{cases}
G_e  = k_t [N-n_t] n +k_r n p \\
         k_t [N-n_t] n = k_n n_t p\\
        n + n_t = p,
\end{cases}
\label{HomEq}
\end{equation}
where $G_e = J_e/L_{cr}$. The system (\ref{HomEq}) was analyzed in detail in Ref. \onlinecite{Kiligaridis2021}. It is shown that at low excitation intensities, the charge carrier densities $k_r n p\ll k_t N n$, $n_t\ll N$.
Thus, we can neglect the corresponding terms in the system (\ref{HomEq}):

\begin{equation}
\begin{cases}
        G_e  = k_t N n  \\
        k_t N n = k_n n_t p\\
        n_t = p - n.
\end{cases}
\label{HomEq1}
\end{equation}
The joint solution to the system (\ref{HomEq1}) is as follows:

\begin{equation}
\begin{cases}
n=G_e/k_tN\\
p=\left[n + \sqrt{n^2+4G_e/k_n}\right]/2\\
n_t=p-n.
\end{cases}
\label{SolHomEq}
\end{equation}

The photoluminescence emission power is as follows:
\begin{equation}
    PL =\xi \hbar \bar{\omega} \frac{k_r G_e^2} {2 k_t^2 N^2} \left[1+\sqrt{1+\frac{4(k_t N)^2}{k_n G_e}} \right] L_{cr} S_{cr},
    \label{HomPL}
\end{equation}
where $\xi = \int F(\vec{r}) d^3r$. Eqs. (\ref{SolHomEq}-\ref{HomPL}) have two possible limits. The first case $\sqrt{k_n G_e} \gg k_t N$ corresponds to the ABC regime, according to the classification proposed in Ref. \onlinecite{Kiligaridis2021}. If the conditions of this regime are met then $p = n\gg n_t$. The PL emission power is the following:

\begin{equation}
        PL = \xi \hbar \bar{\omega} \frac{k_r G_e^2} {k_t^2 N^2} L_{cr} S_{cr}.
\end{equation}

The second case $\sqrt{k_n G_e} \ll k_t N$ corresponds to the SRH regime. The carrier densities are the following:
\begin{equation}
\begin{cases}
    n=G_e/k_t N\\
    p=n_t = \sqrt{G_e/k_n}.
\end{cases}
\end{equation}
The PL emission power is as follows:

\begin{equation}
        PL = \xi \hbar \bar{\omega} \frac{k_r G_e^{3/2}}{k_t N \sqrt{k_n}} L_{cr} S_{cr}.
        \label{SRHHomPL}
\end{equation}

Note that the radiation power in the equation (\ref{SRHHomPL}) is inversely proportional to the trap density $N$, similar to the Stern-Volmer relation (\ref{Stern}). However, the general solution (\ref{HomPL}) has a more complex dependence on the trap density.

\subsection{Axial light propagation}

\begin{figure}[h]
\includegraphics[width=1\linewidth]{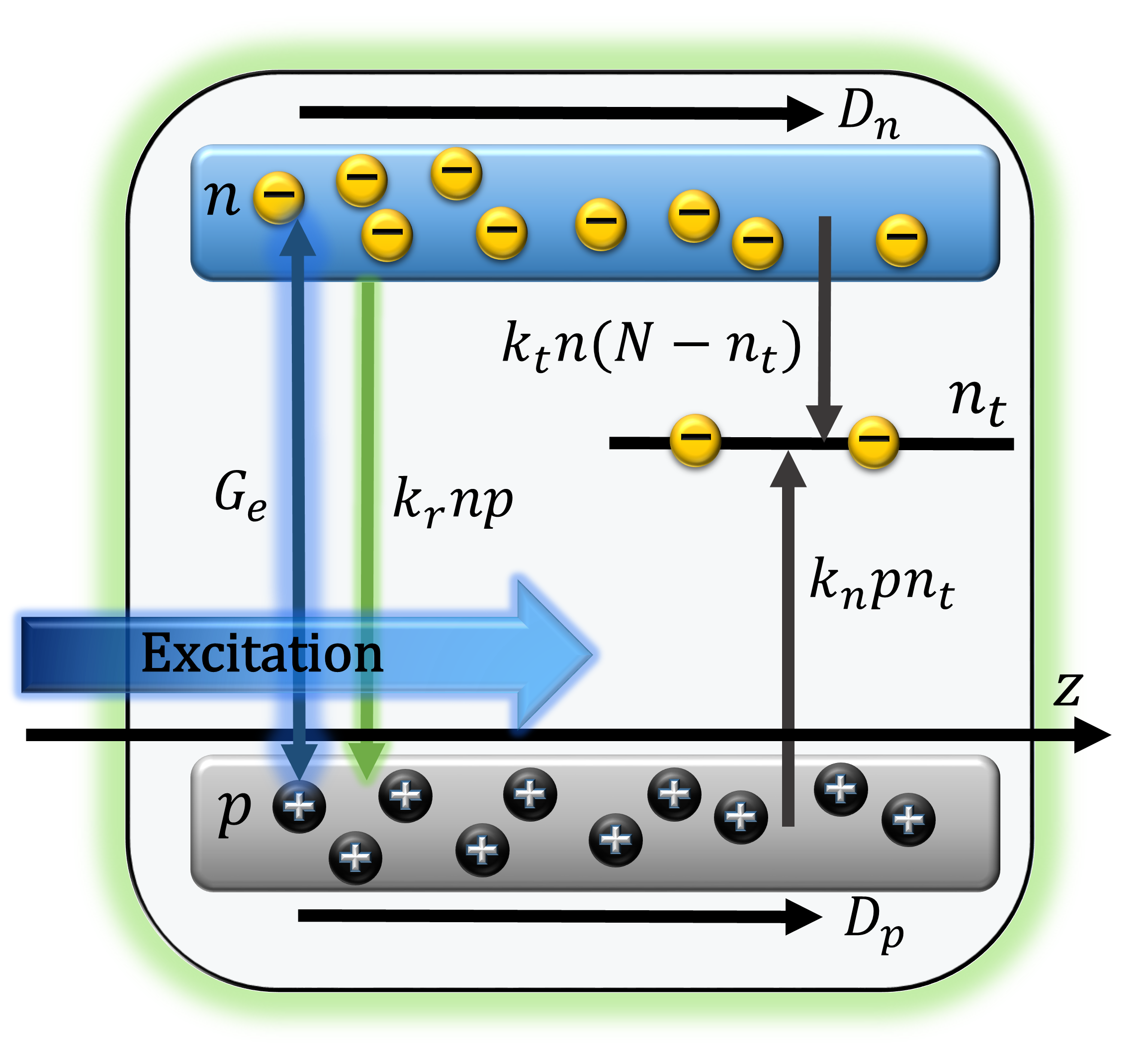}
\centering
\caption{ The schematic picture of the model  for axial excitation propagation.}
\label{AxialPic}
\end{figure}

We assume the shape of the crystal to be an ideal cube. One can consider the propagation of exciting light along the $z$ axis, which is perpendicular to the edge of the crystal, so that the width of the exciting beam is much greater than the transverse area of the crystal $S_{cr}$.
In this case, the carrier and trap density depends only on $z$. The kinetic scheme is shown in Fig. \ref{AxialPic}.
The system (\ref{SRHsimple}) can be written as follows:

\begin{equation}
\begin{cases}
       D_n n''(z) +e\mu_n [n \varphi']'  \\
      \qquad = k_t [N(z)-n_t(z)] n (z)+k_r n(z)p(z) - G_e(z) \\
       D_p   p''(z) - e\mu_p [p \varphi']' \\
       \qquad= k_n n_t(z) p(z) (z)+k_r n(z)p(z) - G_e(z) \\
       k_n n_t(z) p (z)=k_t [N(z)-n_t(z)] n (z) \\
       \varepsilon \varepsilon_0 \varphi''(z) = -e [p(z) - n(z) -n_t(z)],
\end{cases}
\label{SRHaxialField}
\end{equation}
where the volume generation rate is described by the Beer–Lambert law in approximation of the total absorption of excitation light:

\begin{equation}
    G_e(z) = \frac{I_e}{\hbar\omega_e l_e} \exp(-z/l_e) \equiv J_e/l_e\exp(-z/l_e),
\end{equation}
where $l_e$ is the excitation light extinction length. The system obeys the following boundary conditions:

\begin{eqnarray}
   &&n'(0) = p'(0) = 0, \nonumber\\
  &&n'(L_{cr})=   p'(L_{cr}) = 0, \nonumber\\
   &&\varphi'(0) = \varphi'(L_{cr}) = 0,
\end{eqnarray}
where $L_{cr}$ is the longitudinal length of the crystal. The charge conservation law equation (\ref{Charge}) can be rewritten as:

\begin{equation}
    \int\limits_0^{L_{cr}} [n(z) + n_t(z)] dz = \int\limits_0^{L_{cr}} p(z) dz.
    \label{Chargez}
\end{equation}

 The measured PL emission intensity is as follows:
\begin{equation}
    PL=\hbar \bar{\omega} k_r \int\limits_0^{L_{cr}} n(z) p(z) F(z) dz.
    \label{PLaxial}
\end{equation}

In the experiment \cite{TianAdvMat2023} the radiation emitted in the opposite direction relative to the laser beam was measured. Therefore, we use a simple approximation of the radiation survival probability function $F(z) = \exp(-\delta z)$, which takes into account reabsorption.

Let's consider a special case of this model (\ref{SRHaxialField}) assuming  that the diffusion coefficients of electrons and holes are equal and $k_n \gg k_t, k_r$.
This case corresponds to the ABC-like regime.
In this case, the charge density included on the right side of the equation for the potential is quite small, and the resulting field can be neglected. In the results section, we check the applicability of this approximation. Thus the system can be further simplified:

\begin{equation}
\begin{cases}
       D_n n''(z)  = k_t [N(z)-n_t(z)] n (z)+k_r n(z)p(z) - G_e(z) \\
       D_p   p''(z)  = k_n n_t(z) p(z) (z)+k_r n(z)p(z) - G_e(z) \\
       k_n n_t(z) p (z)=k_t [N(z)-n_t(z)] n (z).
\end{cases}
\label{SRHaxial}
\end{equation}

The system (\ref{SRHaxial})  allows us to relate the densities of electrons and holes as follows:

\begin{equation}
    D_n n''(z)= D_p  p''(z).
\end{equation}

Integrating this equation twice with respect to the boundary conditions, we can obtain the following ratio:

\begin{equation}
    D_n[n(z)-n(0)]=D_p [p(z)-p(0)]
\end{equation}
from which follows:

\begin{equation}
    p(z)=\frac{D_n}{D_p} n(z) + \tilde{p}
    \label{ElHole}
\end{equation}
and
\begin{equation}
    n_t(z)=\frac{k_t N(z)}{k_n p(z)+k_t n(z)}.
    \label{Nt}
\end{equation}
The constant parameter $\tilde{p}$  can be found using the charge conservation law (\ref{Chargez})
\begin{eqnarray}
\int\limits_0^{L_{cr}} n(z) \left[1  + \frac{k_t N(z)}{k_n (D_n/D_p n(z)+\tilde{p})+k_n n(z) }\right] dz\qquad \nonumber\\
= \int\limits_0^{L_{cr}} \left[\frac{D_n}{D_p} n(z) + \tilde{p}\right] dz.
\label{Chargep}
\end{eqnarray}

The size of the spatial distribution $n(z)$ in Eqs.(\ref{SRHaxial}) is characterized by the diffusion length:

\begin{equation}
    L_d = \sqrt{\frac{D_n}{k_t N}},
\end{equation}
where $N$ is a characteristic defect density. As mentioned in the introduction, the estimated extinction length of the excitation light is much shorter than the diffusion length. When $l_e \ll L_d$, the system of equations (\ref{SRHaxial}) can be rewritten as:

\begin{equation}
\begin{cases}
       D_n n''(z)  = k_t [N(z)-n_t(z)] n (z)+k_r n(z)p(z) \\
       D_p   p''(z) = k_n n_t(z) p(z) (z)+k_r n(z)p(z)  \\
       k_n n_t(z) p (z)=k_t [N(z)-n_t(z)] n (z)
\end{cases}
\label{SRHaxialSimple}
\end{equation}
with the following boundary conditions:

\begin{equation}
-D_n n'(0) =-D_p  p'(0) = J_e,  \quad    n'(L_{cr})=   p'(L_{cr}) = 0.
\label{nBoundary}
\end{equation}

It follows from relations (\ref{ElHole}-\ref{Nt}) that only the first equation in system (\ref{SRHaxial}) needs to be solved to obtain a solution to the system. The same is true for system (\ref{SRHaxialSimple}).
These equations do not have an analytical solution to an arbitrarily spatial distribution of the trap density $N(z)$. However, an analytic solution to these equations in systems (\ref{SRHaxial}) and (\ref{SRHaxialSimple}) can be found in the case of a uniform trap profile. This solution is provided in supplementary material 1.

At low excitation power, one can neglect the terms of the radiative recombination in the system (\ref{SRHaxial}) and not take into account the filling of traps, as was done in the rapid diffusion limit during the transition from system (\ref{HomEq}) to (\ref{HomEq1}). Finally, we obtain:
\begin{equation}
\begin{cases}
         D_n n''(z)  = k_t N(z) n (z) - G_e(z) \\
       D_p   p''(z) = k_n n_t(z) p(z) (z) - G_e(z) \\
       k_n n_t(z) p (z)=k_t N(z) n (z).
\end{cases}
\label{SRHspec}
\end{equation}
System (\ref{SRHspec}) will be used for calculations and for the construction of analytical approximations.

\section{Photoinduced trap dynamics}

In the original work \cite{TianAdvMat2023}, the authors suggested that the photoinduced creation of recombination centers under the influence of the control light is responsible for luminescence quenching.
In our model, we also assume that the main trap creation mechanism in the band gap is the photoinduced activation of defects. We also assume the presence of stationary traps that are not affected by light.

\begin{equation}
    N(z,t)=N_{st} + N_A(z,t),
\end{equation}
where $N_{st}$ is the density of the stationary traps and $N_A(z,t)$ is the density of the active traps, the number of which is affected by the influence of the control light. We will consider the spatial distribution of stationary traps to be homogeneous.

The mechanism of photoinduced trap creation is currently not quite clear. The possible mechanisms will be covered in detail in the Discussion. Nevertheless, experiments show that, in the absence of the exciting light, the control light affects the defects directly; in the absence of both the control and excitation light, the traps decay by themselves very slowly.
The kinetic equation for the trap density can be written as follows:

\begin{eqnarray}
    \partial_t N_A(z,t) = \sigma_c(\omega_c,T) J_c e^{-z/l_c} \left[N_M -N_A \right]\qquad\nonumber\\ -\gamma_A N_A + \gamma_{cr} [N_M - N_A],
\end{eqnarray}
where
$$\sigma_c(\omega_c,T) \equiv \sigma_{0} (\omega_c) \exp\left(-\frac{\Delta E}{k_B T}\right)$$
 is the effective absorption cross-section of the defect that causes the trap creation under the control light at a given temperature $T$ and control light frequency $\omega_c$, where $N_M$ is the maximal number of active defects, $l_c$ is the control light extinction length, and $\gamma_{cr}$ and $\gamma_{A}$ are the corresponding rates of spontaneous trap creation and self-destruction. The presence of the activation barrier is explained in the Discussion. Here we assume that the control and excitation light enter the same surface.

Since the experimentally observed characteristic time of spontaneous self-destruction of the traps occurs on the scale of hours, we neglect this process in future considerations.
We also assume that the excitation light generates the traps. By considering this effect and neglecting the spontaneous self-destruction processes, one can get the following equation:

\begin{equation}
    \partial_t N_A(z,t) = K_{cr} (z) \left[N_M - N_A(z,t)\right]- R_A.
    \label{DefCreation}
\end{equation}
Here $R_A$ is the trap's annihilation rate and $K_{cr} (z)$  is the effective creation rate constant:
$$K_{cr} (z) \equiv \sigma_c (\omega_c) J_c e^{-z/l_c} + \sigma_e (\omega_e) J_e e^{-z/l_e}.$$
where
$$\sigma_e(\omega_e,T) \equiv \sigma_{0} (\omega_e) \exp\left(-\frac{\Delta E}{k_B T}\right)$$  is effective absorption cross-section of the defect that causes the trap creation under the excitation light.

In the limit $l_e \ll L_d$ $K_{cr}$ has the form:
 $$K_{cr} (z) = \sigma_c (\omega_c) J_c e^{-z/l_c}. $$

In the original work, the authors clearly demonstrated that the excitation light is responsible for the restoration of luminescence. We have already shown that the excitation light's extinction length $l_e$ is extremely short and amounts to hundreds of nanometers for most of the experiments carried out; therefore, the excitation light will not be able to destroy all of the generated traps, since their minimum depth is determined by the extinction length of the control light $l_c$, which exceeds the extinction length of the exciting light by at least one order of magnitude.

\begin{figure}[h]
\includegraphics[width=1\linewidth]{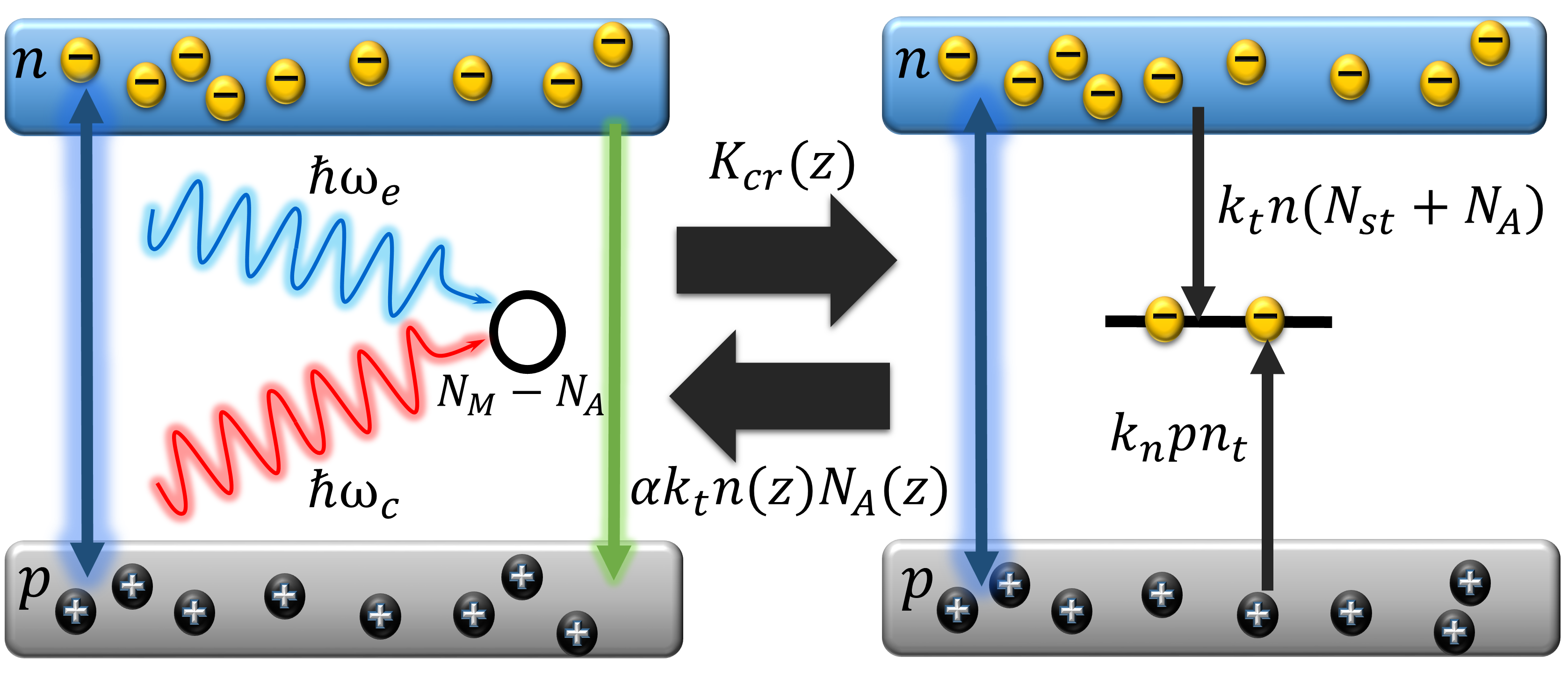}
\centering
\caption{Schematic picture of the trap kinetics model.}
\label{PhotoSwitchingShort}
\end{figure}

Thus, it becomes obvious that, although the excitation light is responsible for the destruction of traps, it does not act through direct absorption, but does so through some intermediate process. In our model, all of the excitation light is spent on creating new traps in a narrow zone near the crystal surface as well as on generating charge carriers. Thus, the most likely mechanism is the interaction of charge carriers with traps, which leads to the annihilation of the latter. We consider some possible processes that may be responsible for trap annihilation in the Discussion.

In our model, annihilation of an active trap can occur as a result of the recombination of an electron trapped in it with a hole.
The probability of the trap being destroyed as a result of such an event is $\alpha$.
Thus, the annihilation  rate can be written as:
$$R_A=\alpha k_t n(z,t) N_A(z,t).$$
Finally, we have the following equation:
\begin{equation}
    \partial_t N_A(z,t) = K_{cr}(z) \left[N_M - N_A(z,t)\right] - \alpha k_t n(z,t) N_A(z,t).
    \label{DefDyn}
\end{equation}

The characteristic times of the trap annihilation and creation far exceed the characteristic times of the electronic subsystem kinetics. Thus, solving the system (\ref{SRHspec}) we can consider $N_A$ as constant. The schematic picture of the trap kinetic model is presented in Fig. \ref{PhotoSwitchingShort}.

\section{Results}
\subsection{Homogeneous limit}

When the condition of the rapid diffusion limit is applied and the extinction length of the control light far exceeds the lateral length of the crystal,
the distributions of the charge carriers and defects are homogeneous. Therefore, the following equation for the trap density is applied:

\begin{equation}
    d_t N_A = \sigma_c J_c [N_M - N_A(t)] - \alpha k_t n(t) N_A(t).
    \label{NAA}
\end{equation}
The charge recombination kinetics is described by the system (\ref{HomEq1}). From the first equation we have $$n=G_e/k_t (N_{st}+N_A).$$
Thus, the equation (\ref{NAA}) can be rewritten in the following form:
\begin{equation}
    d_t N_A = \sigma_c J_c [N_M - N_A(t)] - \frac{\alpha G_e}{k_t [N_{st}+N_A(t)]} N_A(t).
    \label{N_A}
\end{equation}
The photoluminescence emission power in this limit is given by Eq.(\ref{HomPL}):
\begin{equation}
    PL(t) = \frac{\xi \hbar \bar{\omega} L_{cr} S_{cr} k_r G_e^2} {2 k_t^2 (N_A(t)+N_{st})^2} \left[1+\sqrt{1+\frac{4 k_t^2 (N_A(t)+N_{st})^2}{k_n G_e}} \right].
    \label{PLN}
\end{equation}

\begin{figure}[h]
\includegraphics[width=1\linewidth]{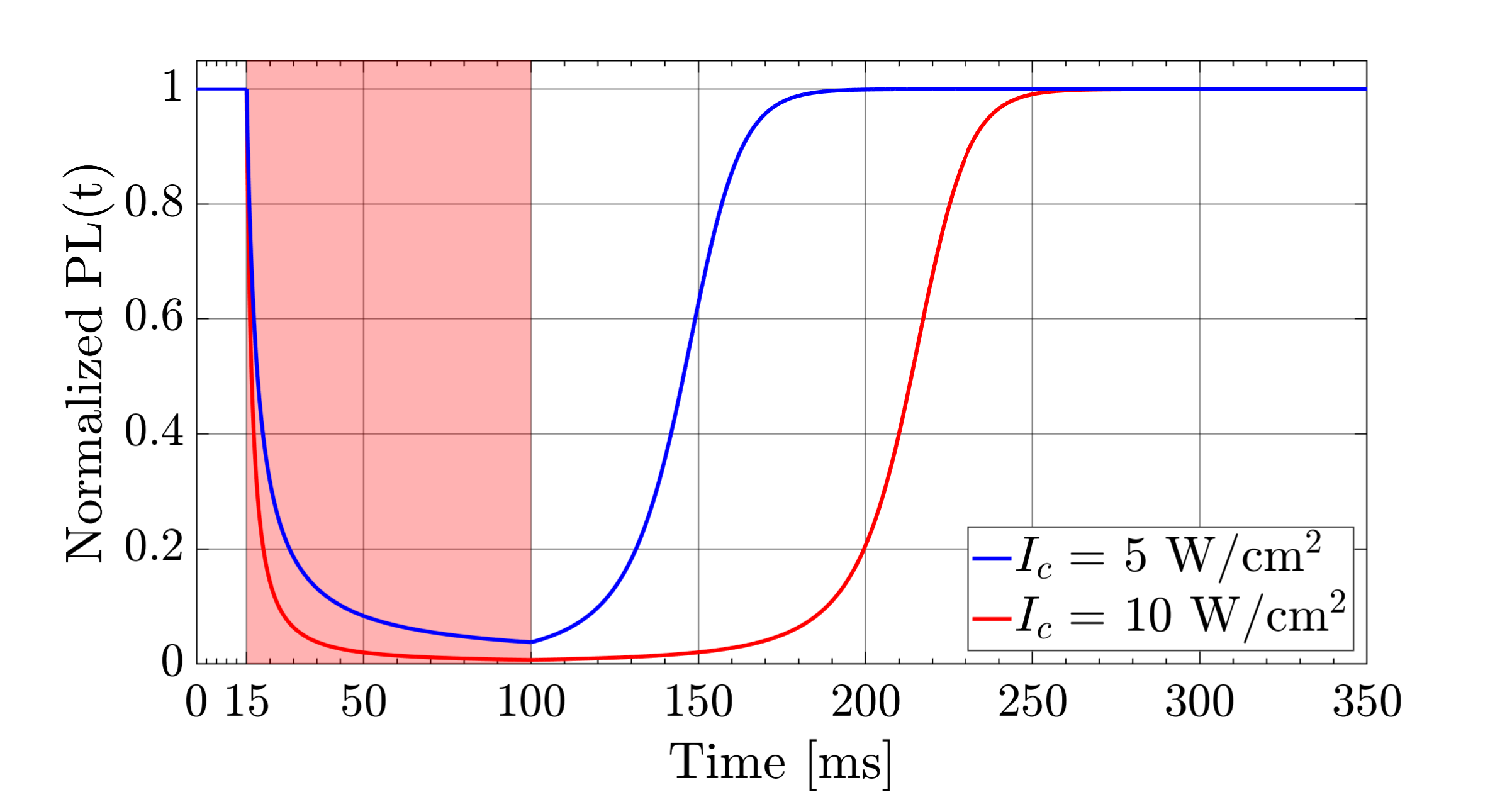}
\centering
\caption{The photoluminescence intensity time dependence in the homogeneous limit at different control light intensities.}
\label{lambert}
\end{figure}

The analytical solution of the Eq.(\ref{N_A}) is shown in supplementary material 2. The characteristic form of the photoswitching curve obtained from this analytical solution and Eq.(\ref{PLN}) is shown in Fig. \ref{lambert}.

The parameters chosen for this calculation are $N_{st}=5\times10^{15}$cm\textsuperscript{-3}, $I_e=1$ W/cm\textsuperscript{2}, $\lambda_e=450$ nm, $\lambda_c=600$ nm, $N_M=2.5\times10^{17}$cm\textsuperscript{-3}, $\alpha=3\times10^{-4}$, $\sigma_c=2 \times 10^{-19}$cm\textsuperscript{2}, $k_t=2\times10^{-9}$cm\textsuperscript{3}s\textsuperscript{-1}, $k_n=2\times10^{-5}$cm\textsuperscript{3}s\textsuperscript{-1}, $L_{cr}=10$ $\mu$m.

As seen, the $PL(t)$ dependence is very different from the logistic curve (\ref{OFFdep})  within the homogeneous limit.
It is also not possible to explain the dependence of the rising rate on the luminescence switching efficiency.
Thus, although this limit takes into account the small extinction length of the exciting light, it is clearly not suitable for an accurate description of the phenomenon.

\subsection{General case}

Most of the experiments in the original article were performed for crystals that ranged in size from 5 to 10 microns. These values far exceed the maximal electron diffusion length, as well as the extinction length of the control light. Thus, the homogeneous limit is not applicable in this case.

 In the general case when the extinction length of the excitation light is much smaller than the diffusion length, the density of traps is described by Eq. (\ref{DefDyn})
where the quasi-stationary densities of the charge carriers can be found by solving the system (\ref{SRHspec})  with the boundary conditions (\ref{nBoundary}). These equations do not have a general analytical solution. However, some approximate analytical solutions can be applied under certain conditions:

{\bf The Wentzel-Kramers-Brullien} approximation for the quasi-stationary electron density works well at low switching efficiency values.

{\bf The moving boundary} approximation works well in the case of high switching efficiencies and sufficiently small crystals compared to the control light extinction length.

\begin{figure*}[t]
\center{\includegraphics[width=0.9\linewidth]{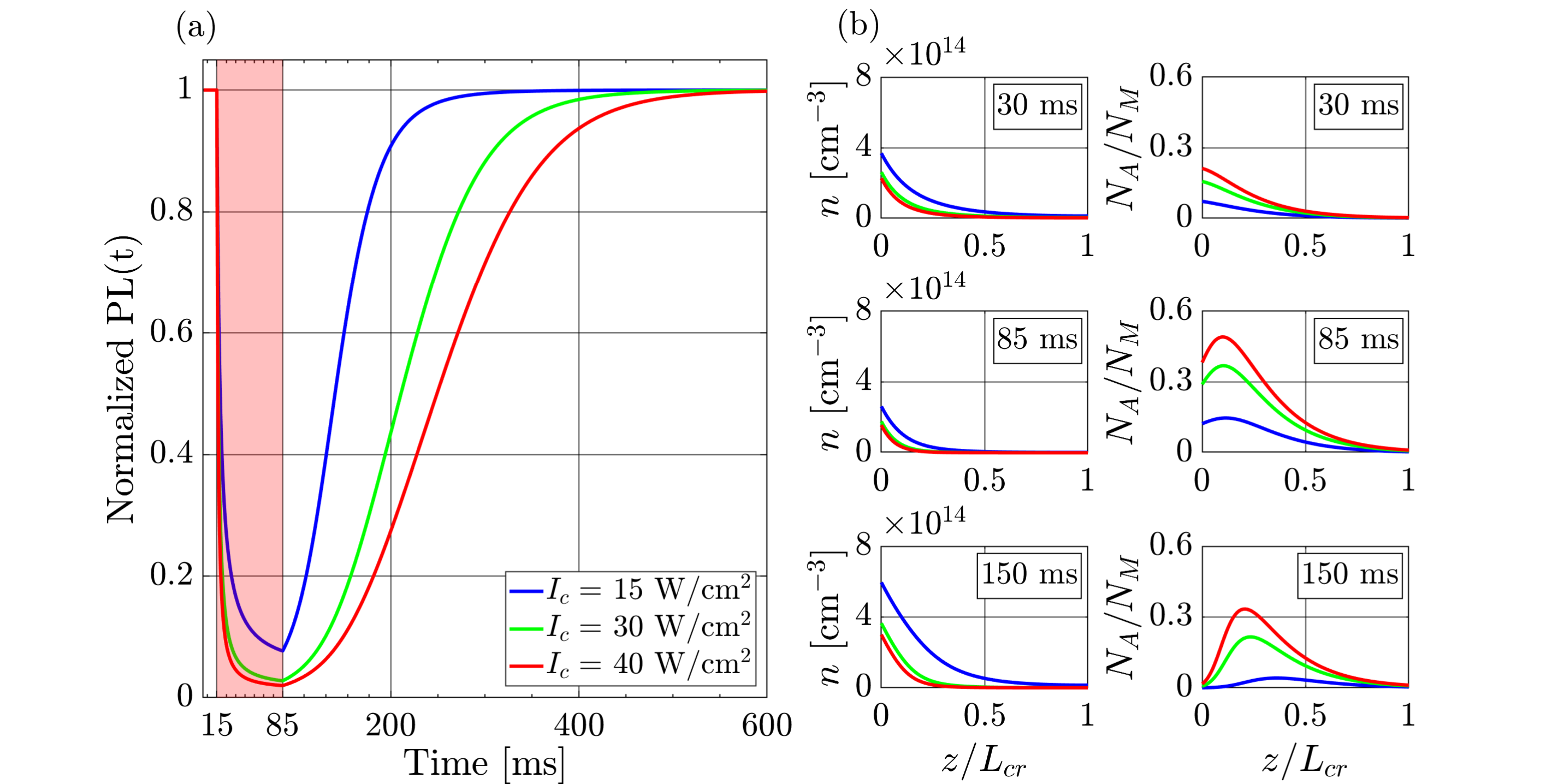}}
\centering
\caption{(a) PL intensity as a function of time at different control light intensities.
 (b) Profiles of the electron density and trap density at different times. The excitation light intensity is 1 W/cm\textsuperscript{2}.}
\label{DifControlPic}
\end{figure*}

 The details of these approximations and the comparison of their predictions with the numerical modeling results are presented in the supplementary material 3.

The results of a numerical solution of Eq.(\ref{DefDyn}) together with the system (\ref{SRHspec}) with the boundary conditions (\ref{nBoundary}) are presented in Figs. (\ref{DifControlPic}-\ref{DiffExc}).

The details of the numerical simulation procedure can be found in supplementary material 5.

Fig. \ref{DifControlPic} shows the photoswitching effect at different intensities of the control light.
It can be seen that the falling rate increases and the rising rate decreases with an increase in the control light intensity.

The parameters of the model are as follows: $N_{st}=5\times10^{15}$cm\textsuperscript{-3}, $\lambda_e=450$ nm, $\lambda_c=600$ nm, $N_M=2.5\times10^{17}$cm\textsuperscript{-3}, $\alpha=1.1\times10^{-4}$, $\sigma_c=2 \times 10^{-19}$cm\textsuperscript{2}, $k_t=2\times10^{-9}$cm\textsuperscript{3}s\textsuperscript{-1}, $k_n=2\times10^{-5}$cm\textsuperscript{3}s\textsuperscript{-1}, $L_{cr}=10$ $\mu$m, $l_c=2$ $\mu$m, $D_n=1$ cm$^2$s$^{-1}$, $D_p=1$ cm$^2$s$^{-1}$. The coefficient $\delta$ is considered equal to $10^{4}$cm\textsuperscript{-1} for all subsequent calculations \cite{Scajev2020}.

We estimated the electric field obtained for the trap profile in Fig. \ref{DifControlPic} for 85 ms at 15 W/cm$^{2}$ control light intensity, which is the most typical intensity used for calculations in this manuscript. The electric potential was calculated using Gauss's theorem:

\begin{equation}
    \varphi(z) = -\frac{e}{\varepsilon \varepsilon_0} \int\limits_0^{z} d\zeta \int\limits_0^{\zeta} [p(\zeta') - n(\zeta') - n_t(\zeta')] d\zeta'.
\end{equation}

\begin{figure}[b]
\includegraphics[width=0.99\linewidth]{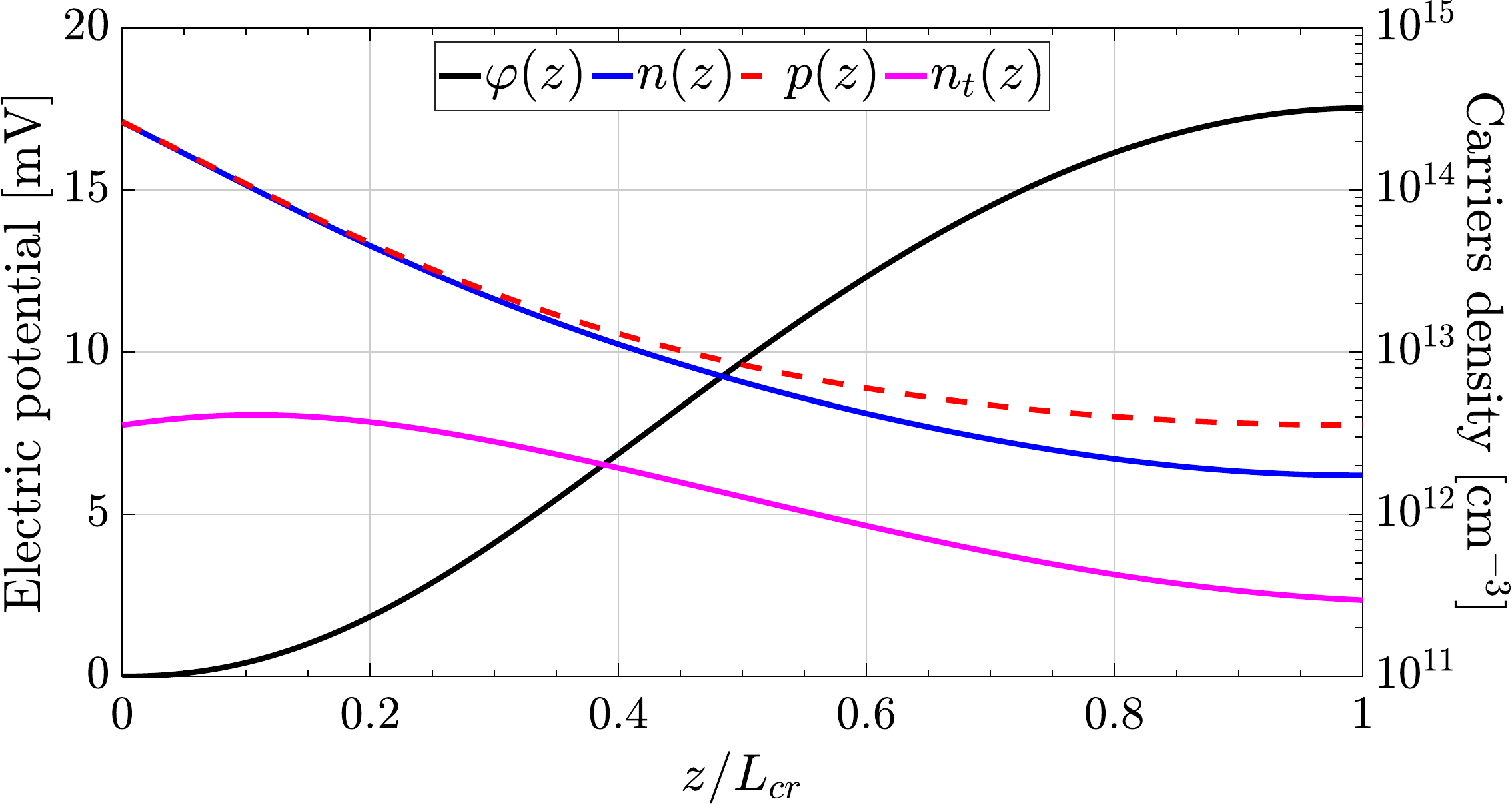}
\centering
\caption{The electric potential distribution within the crystal (black line), electron (blue line), hole (red dashed line) and trapped electron (magenta line) densities.}
\label{Field}
\end{figure}

The value of the static dielectric permittivity of MAPbBr\textsubscript{3} is not clearly defined. There are several report with different values in the range of 20 to 100 \cite{Miyata2017, Govinda2017, Ramya2022}. This variation in values is probably related to the different morphology and nature of the samples. For our calculations, we have chosen an $\varepsilon$ equal to 40. The result of the calculation, as well as the densities of the carriers, are shown in Fig. \ref{Field}. It can be seen that the maximum potential difference corresponds to a value of about 17 mV (the corresponding energy is 17 meV), which is less than the characteristic energy of diffusion motion at room temperature equal to 26 meV, but it is approaching this value. However, the calculation results for two other profiles at the same control intensity give values of a 6 and 1.5 mV electric potential difference for profiles at 30 and 150 ms, respectively. The results for profiles corresponding to other control intensities may differ a few times. The maximal potential value is found at 60 W/cm\textsuperscript{-2} control intensity at 85 ms. It equals 47 mV and is in the order of $k_B T$. Thus, it can be concluded that the influence of the electric field is not significant in the case of the model parameters used.

The theoretical dependencies of the switching efficiency and the luminescence falling rate on the control light intensity are shown in Fig. \ref{SEcontrolPic}. It can be seen that the switching efficiency dependence has a characteristic S-shape form, which corresponds to the experimental observations. The presence of falling rate saturation at low intensities is a result of the effect of excitation light on luminescence quenching by control light. In this case, the next condition is met $\sigma_c J_c N_A \ll \alpha k_t n N_A$, thus the equation describing the traps dynamics changes as follows:

\begin{equation}
 \partial_t N_A = \sigma_c J_c N_M -\alpha k_t n(t,z) N_A(t,z).
\end{equation}
Therefore, the falling rate is determined by the traps annihilation rate. The results of numerical calculations of the same dependence at different excitation intensities are provided in supplementary material 5.

\begin{figure}[t]
\flushleft{(a)}
\begin{minipage}[h]{0.99\linewidth}
\center{\includegraphics[width=1\linewidth]{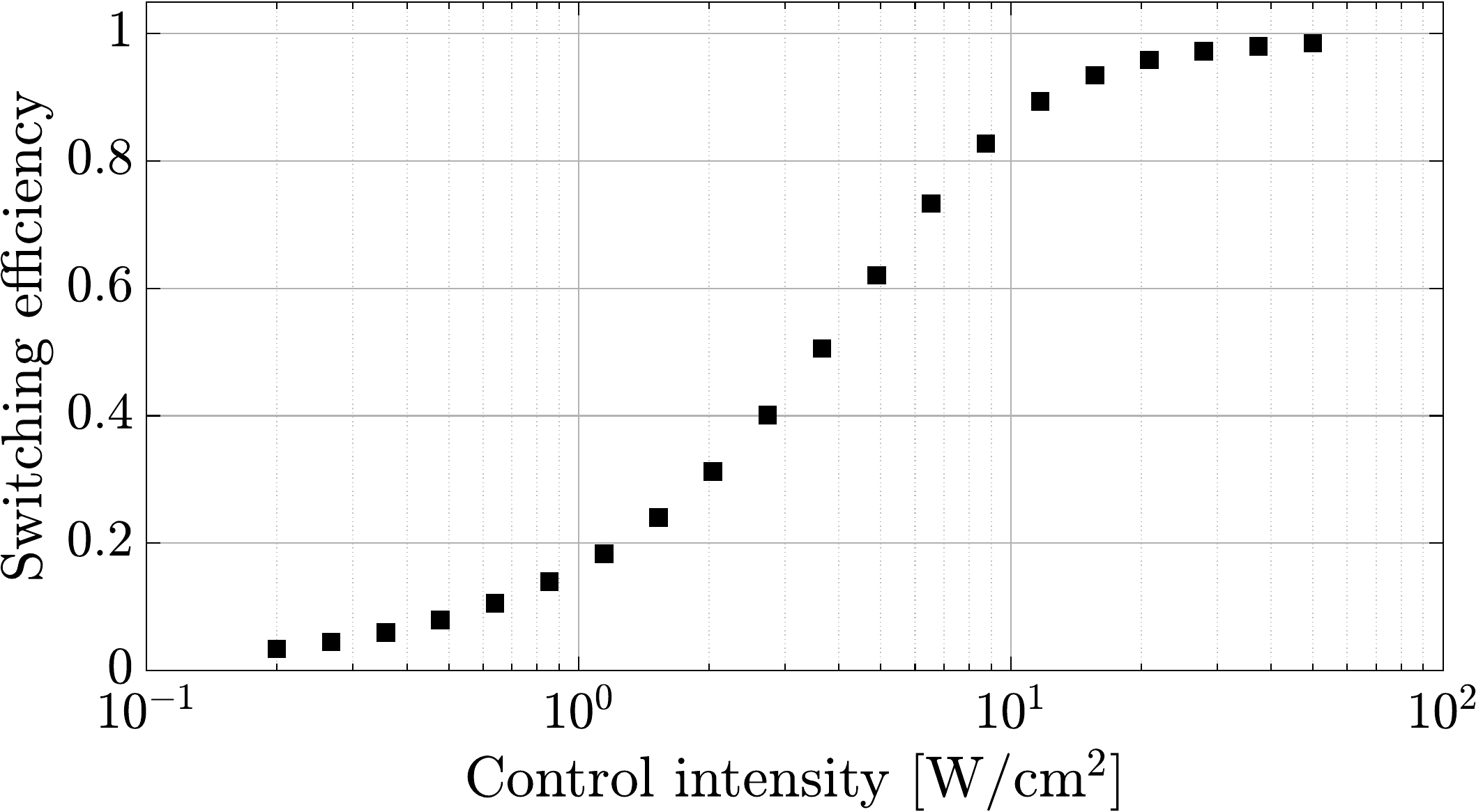}}
\end{minipage}
\vfill
\flushleft{(b)}
\begin{minipage}[h]{0.99\linewidth}
\center{\includegraphics[width=1\linewidth]{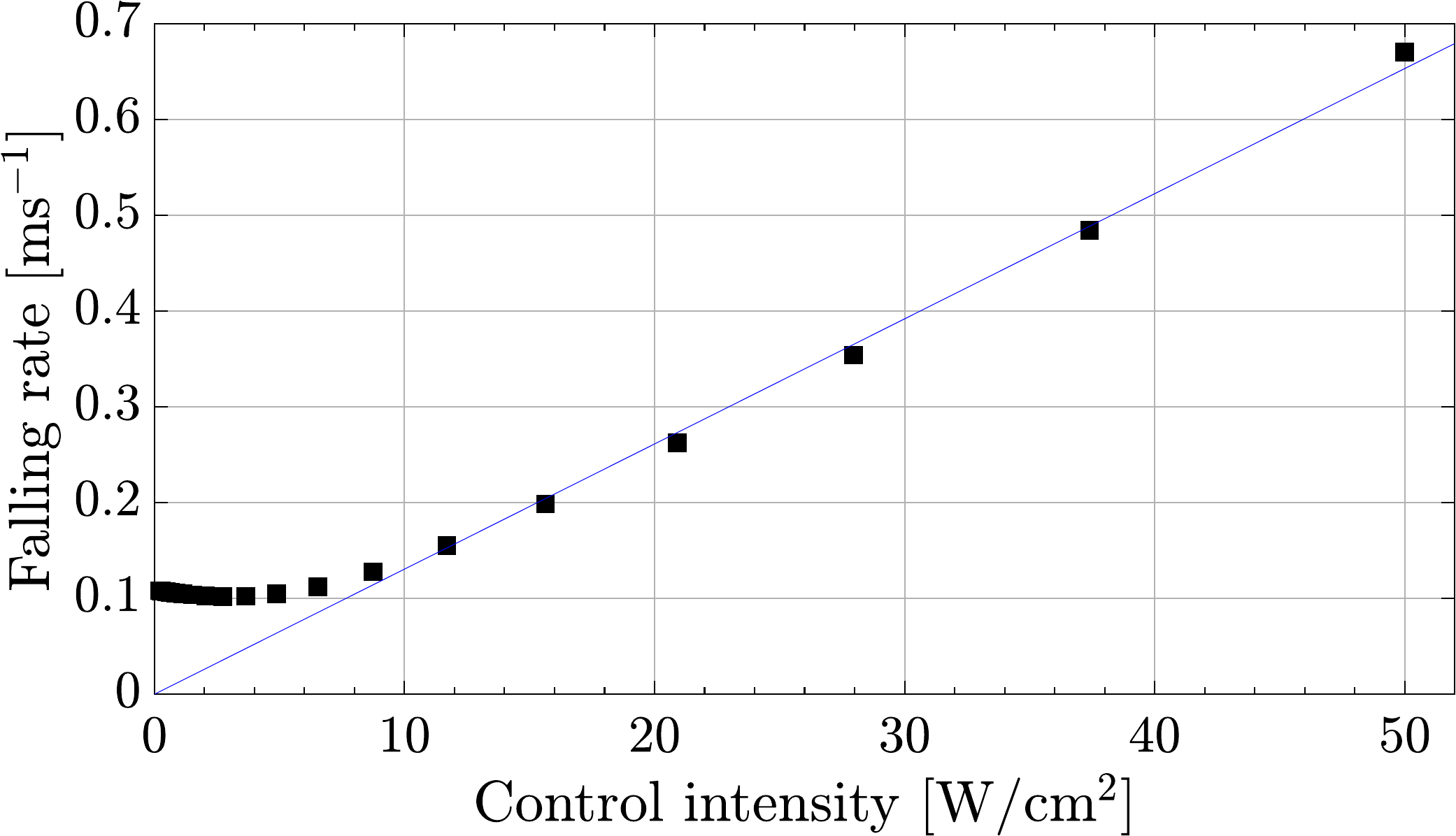}}
\end{minipage}
\caption{(a) The dependence of the switching efficiency on the control light intensity. (b) The luminescence falling rate dependence on the control light intensity. Blue line represents the linear fit for the points corresponding to control intensity greater than 10 W/cm$^2$.}
\label{SEcontrolPic}
\end{figure}

The theoretical dependencies of the switching efficiency and the luminescence rising rate on the excitation light intensity are shown in Fig. \ref{DiffExc}. As seen, switching efficiency decreases with increasing excitation light intensity, initially linearly and then more slowly as it approaches zero. The rising rate is an almost linear function of the excitation light intensity.
That is what is seen in the experiment. The theoretical PL falling and rising curves are also in good agreement with the exponential and logistic curves correspondingly. Details are presented in supplementary material 5.

Thus, the simulation results show that, in the limit of the small extinction length of the excitation light, our model qualitatively explains almost all the properties observed experimentally.

Nevertheless, the model in the limit of small extinction length of excitation light still has one drawback: it does not explain the strong correlation between the excitation light absorption and the photoswitching raising rate observed in the experiment. To explain this, we have to use the model in a more general form (\ref{SRHaxial}) and the complete equation for trap density (\ref{DefDyn}).

\begin{figure}[b]
\flushleft{(a)}
\begin{minipage}[h]{0.99\linewidth}
\center{\includegraphics[width=1\linewidth]{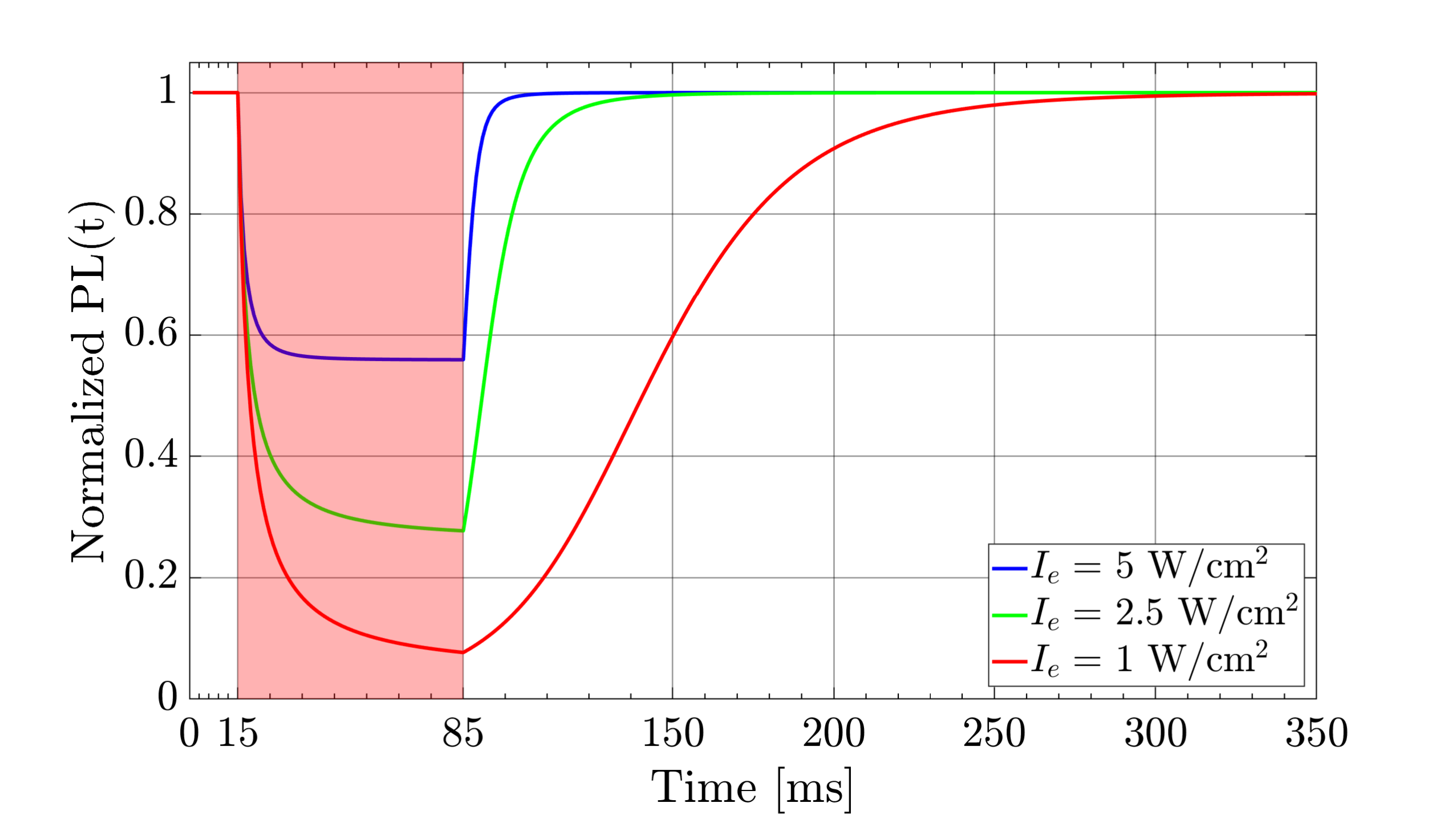}}
\end{minipage}
\vfill
\flushleft{(b)}
\begin{minipage}[h]{0.99\linewidth}
\center{\includegraphics[width=1\linewidth]{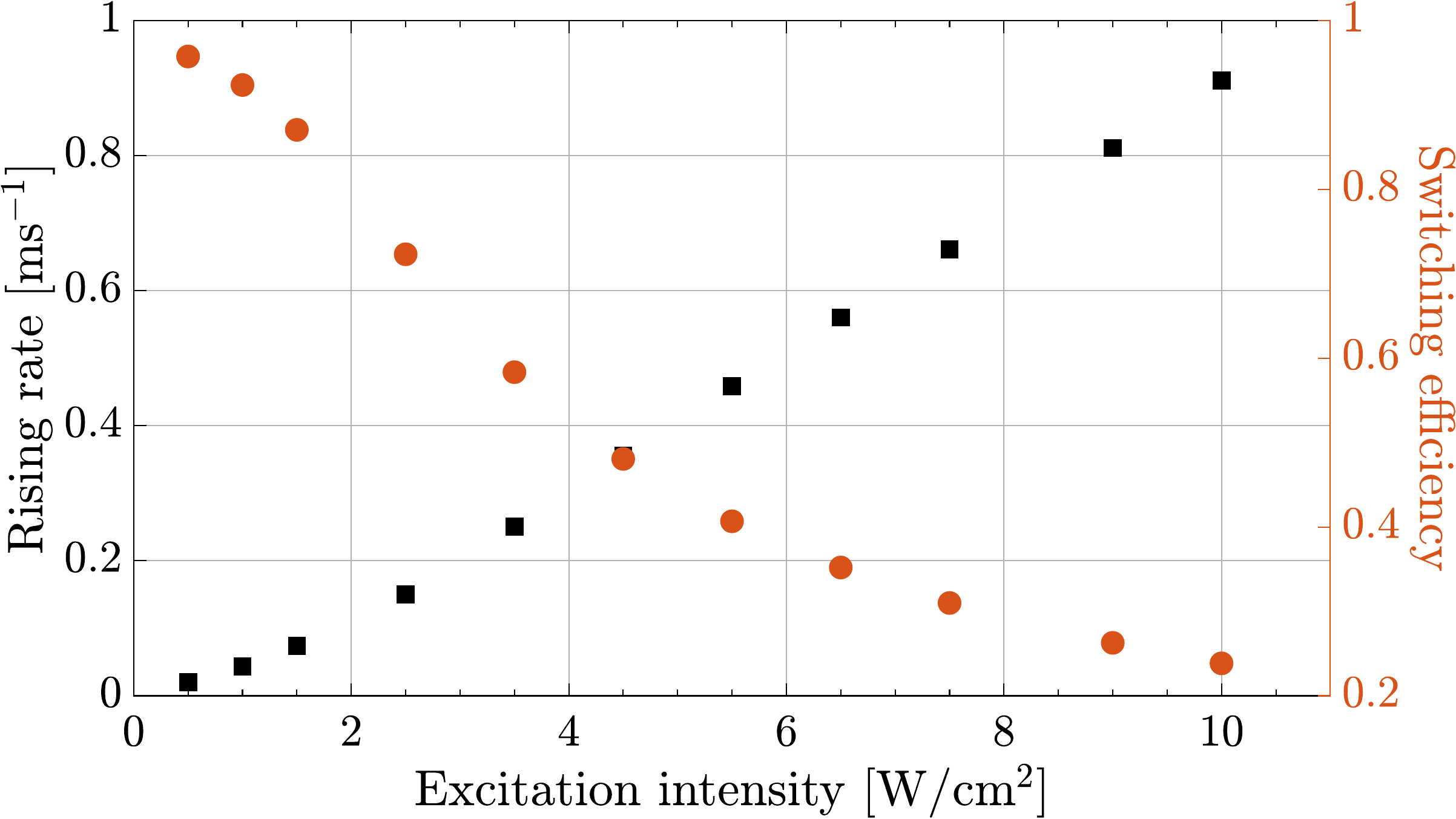}}
\end{minipage}
\caption{ (a) PL intensity as a function of time at different excitation light intensities. (b) The dependence of the switching efficiency (orange circles) and the rising rate (black squares) on the excitation light intensity. The control light intensity is 15 W/cm\textsuperscript{2}.}
\label{DiffExc}
\end{figure}

\begin{figure}[h]
\flushleft{(a)}
\begin{minipage}[h]{0.99\linewidth}
\center{\includegraphics[width=1\linewidth]{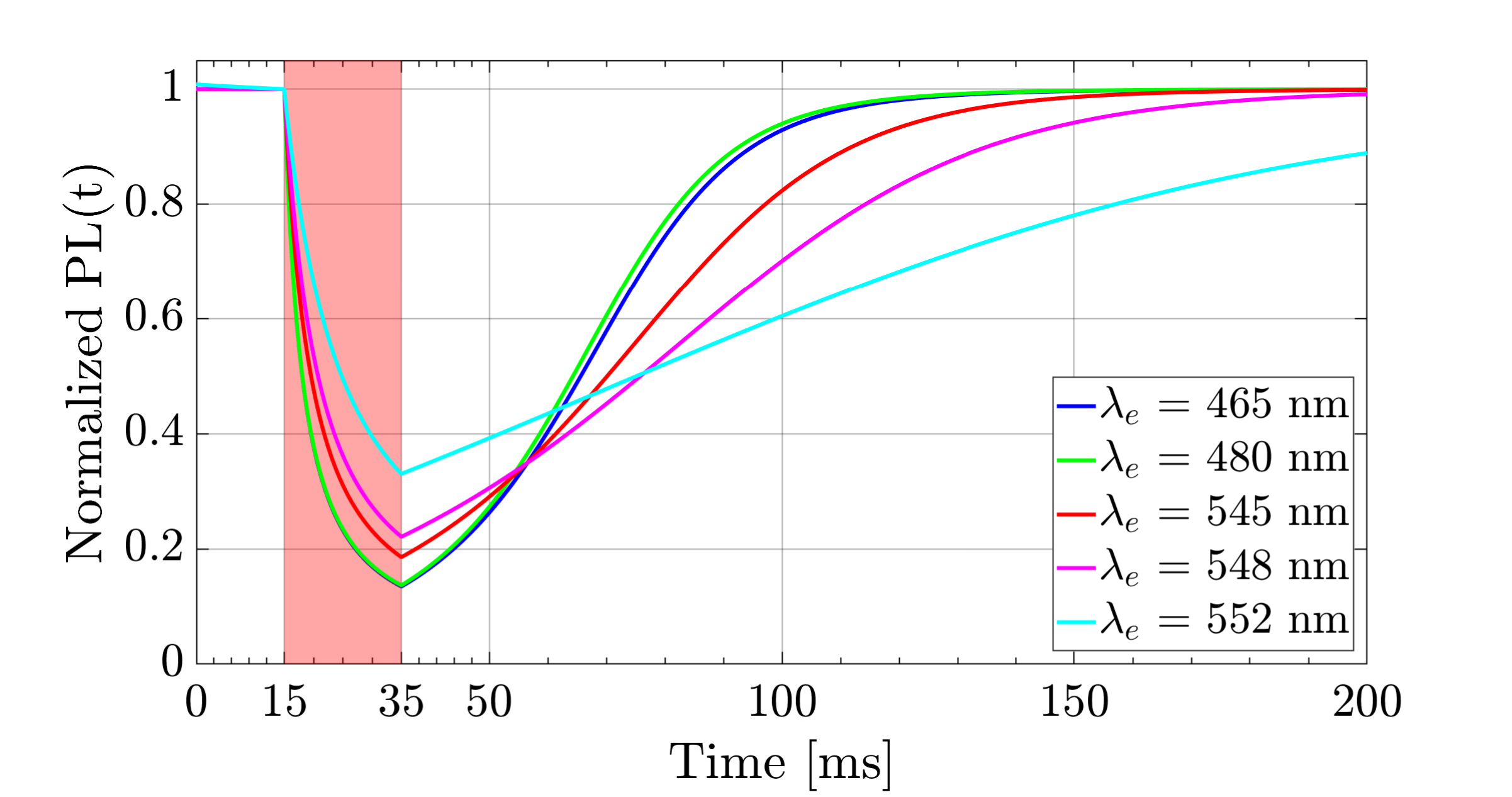}}
\end{minipage}
\vfill
\flushleft{(b)}
\begin{minipage}[h]{0.99\linewidth}
\center{\includegraphics[width=1\linewidth]{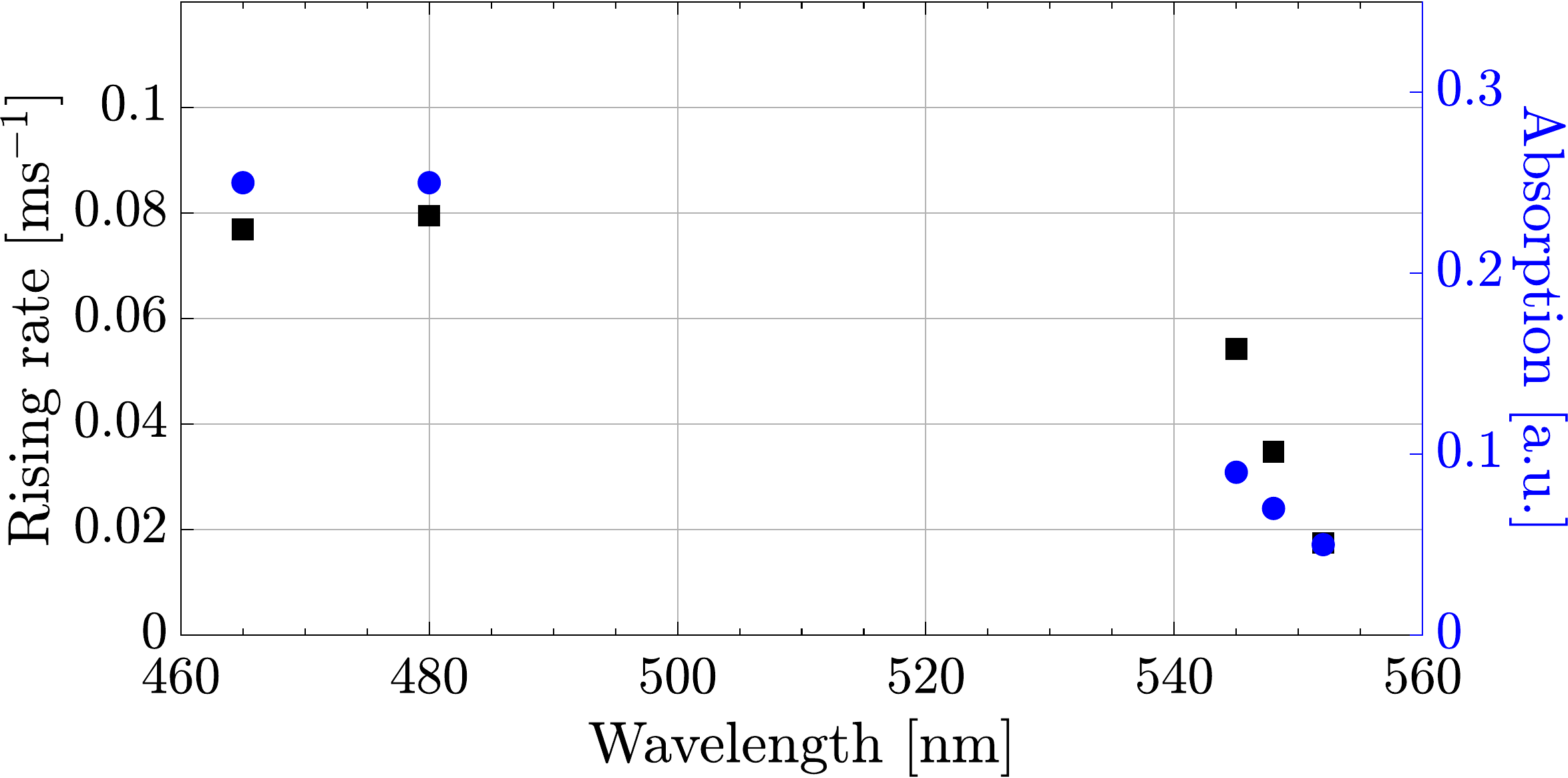}}
\end{minipage}
\caption{(a) PL intensity as a function of time at different excitation light wavelength. (b) The relative absorption of the excitation light (blue circles) and the rising rate (black squares) as the function of the excitation light wavelength. The intensity of the control light is 15 W/cm\textsuperscript{2}. The intensity of the excitation light is 1 W/cm\textsuperscript{2}.}
\label{DifLambda}
\end{figure}

The results of the numerical simulations in this case are shown in Fig. \ref{DifLambda}. It can be seen that there is a correlation between absorption and rising rate due to the different extinction lengths of the excitation light.

The parameters chosen in this case are as follows: $N_{st}=5\times10^{15}$cm\textsuperscript{-3}, $\lambda_e=450$ nm, $\lambda_c=600$ nm, $N_M=2.5\times10^{17}$cm\textsuperscript{-3}, $\alpha=1.92\times10^{-4}$, $\sigma_c=3 \times 10^{-19}$cm\textsuperscript{2}, $\sigma_e=6.3 \times 10^{-18}$cm\textsuperscript{2}, $k_t=10^{-9}$cm\textsuperscript{3}s\textsuperscript{-1}, $k_n=10^{-5}$cm\textsuperscript{3}s\textsuperscript{-1}, $L_{cr}=10$ $\mu$m, $l_c=2$ $\mu$m, $D_n=1$ cm$^2$s$^{-1}$, $D_p=1$ cm$^2$s$^{-1}$. The extinction length of excitation light is calculated using the next relation $l_e (\lambda)/l_e(465$ nm$) =A(465$ nm$)  /A (\lambda)$, where $A(\lambda)$ is the relative absorbance. Here $A(465$ nm$)= 0.25$, $l_e(465$ nm$)=200$ nm. The wavelength values  and absorption data correspond to the experimental absorption spectra \cite{TianAdvMat2023}.

In conclusion, the proposed model is able to explain all the properties of the photoswitching  presented in the original work.

\section{Discussion}
\subsection{Nature of the traps}

Great attention has been devoted to the study of defects in perovskite materials. Due to a variety of simulations, including calculations based on the density functional method and molecular dynamics \cite{Mannodi2020, Mosconi2016, Qiao2020,  Meggiolaro2019} as well as experimental efforts \cite{Jin2020, Hong2018, Cho2021, Srivastava2023, Chen2020, MERDASA2020116916},  it is known that almost any point defect, including interstices, vacancies, antisites of halides, anions, and lead, creates an additional energy level. However, for MAPbBr\textsubscript{3}, which is the main focus of this article, the most likely candidates for deep trap formation are lead interstices and halide vacancies \cite{Xue2021}. The fact that there was a large excess of PbBr\textsubscript{2} in the system studied during synthesis supports the latter statement. We assume that the creation of trap states corresponds to the transition of an atom from a lattice node to an interstitial space, generating a Frenkel defect, \textit{i.e.} a vacancy-interstitial pair.
Because two characteristic processes of changing the luminescence emission power were observed in the experiment, namely the fast photoswitching and the slow luminescence intensity enhancement, it can be assumed  that one of them corresponds to the formation of traps associated with halides and the other with lead. Stationary traps can correspond to defects of another type or be unpaired interstices/vacancies, \textit{i.e.} Schottky defects. However, it is unknown what kind of traps are responsible for the relatively fast photoswiching process.

In the experiment, the creation of defects occurs mostly under a red control light. Because the energy of direct light absorption by a single atom lies in the UV range, such a mechanism cannot be responsible for the creation of new defect states. Long before the creation of traps by light excitation was found in perovskite systems, it was studied for other types of solid systems (for example, see Ref. \onlinecite{Klinger1985} and the references therein).
 Quite a number of different mechanisms for the creation and annihilation of deep trap states are known. Nevertheless, all such processes require the transfer of the energy of the excited local electronic subsystem to some defect of the crystal structure. The possible candidates for such a defect are a conglomeration of several shallow traps, a single local crystal structure defect or a group of them.
 We assume that light with photon energy below and above the band gap is able to excite these localized electronic subsystems.  The nature of these localized excitations remains unclear. It is possible, however, that these excitations are connected either by the candidate defect itself or indirectly associated with it through self-trapped (small) polarons \cite{Carpenella2023, Wong2020, Buizza2021} or self-trapped excitons (which are more often observed for all-inorganic perovskites \cite{Pan2022, Wu2015, Li2019}).

\begin{figure}[t]
\includegraphics[width=1\linewidth]{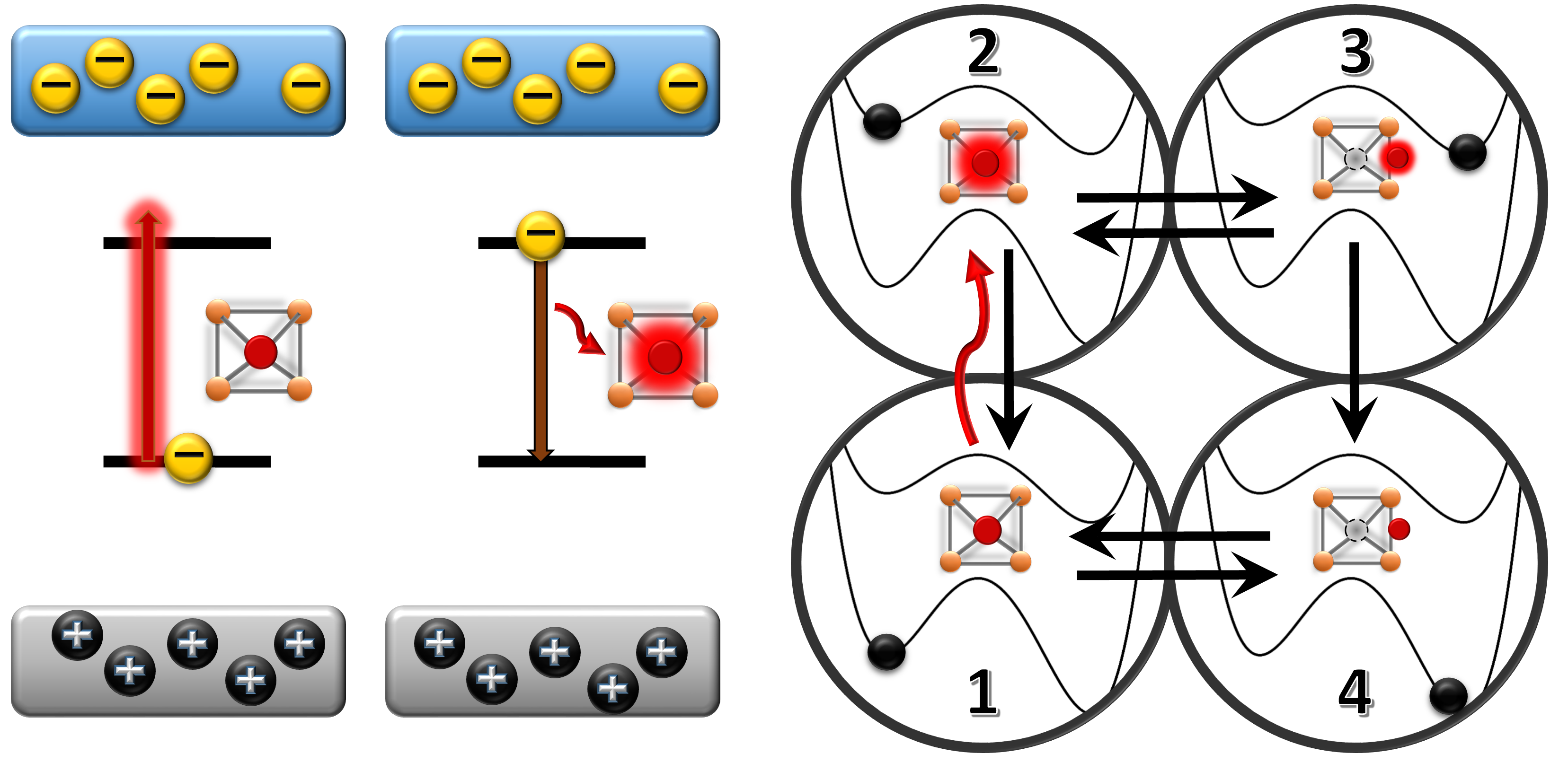}
\centering
\caption{The schematic picture of the deep trap state creation mechanism.}
\label{PhotoSwitching}
\end{figure}

To summarize, we can propose a phenomenological model for the creation of a light-induced deep trap state, shown in Fig. \ref{PhotoSwitching}.

The figure shows an intraband transition under a control light, which is associated with a localized electronic subsystem excitation.
This transition can be ensured by the excitation light as well.
 The energy transfer process transmits the excitation from the localized electronic subsystem to the local defect exciting it (transition 1-2).
 The electronically excited defect can turn from the initial configuration into a final configuration via transition 2-3, or return to the ground state through transition 2-1.
 The excited defect within the final configuration can return back via transition 3-2, or relax to the final configuration ground state via transition 3-4.
 The ground state of the final configuration corresponds to the deep trap state.
 We suggest that the activation barrier for the transition 3-2 is much higher than for transition 2-3.
 In this case, the  relaxation to the final configuration ground state is the most likely.
 Transitions 1-4 and 4-1 are spontaneous processes of trap creation and annihilation and have already been mentioned in previous sections.
 It seems that transitions 2-1 and 3-4 are much faster than the transitions 1-2 and 2-3. So, processes 1-2 and 2-3 are limiting, and the dynamics of trap formation can be described by Eq. (\ref{DefCreation}).

Let's  consider the dynamics of trap annihilation. This effect is even more complex than the formation of traps, as it involves free charges.
 We have considered several possible phenomenological decay mechanisms:

1. A first-order self-destruction of the traps in which the electron from the conduction band is captured. In this mechanism, the rate of the traps annihilation has the following form:
\begin{equation}
    R_A= -k_S n_t(z,t)\frac {N_A(z,t)}{N_{st}+N_A(z,t)} \equiv -k_S n_t^{(A)} (z,t),
    \label{k_S}
\end{equation}
where $k_S$ is the rate constant of filled trap self-destruction.

2. The annihilation of the trap within a non-radiative recombination of a trapped electron with the free hole with some probability $\alpha$. In this case, the following equation can be obtained:
\begin{equation}
    R_A = -\alpha k_t n(z) N_A(z,t),
\end{equation}

Similar mechanism of trap annihilation has already been successfully used to explain the photoluminescence intensity rising in perovskite MAPI films\cite{Fu2016}.

3. The annihilation process that involves multiple traps, where $R_A$ has the form $n_t^{(A)} N_A$, $n N N_A$, $n N_A^2$, etc. This corresponds to the interaction of the defects with each other in the presence of a charge carrier.

The results of the numerical simulations of the Eq.(\ref{DefDyn}) and Eq.(\ref{SRHsimple}) with the first mechanism  Eq.(\ref{k_S}) do not reproduce the control light intensity dependence of the rising rate.
The simulations within the third mechanism do not reproduce the PL rising curve. So, it was realized that the first and the third mechanisms are not suitable to describe the photoswitching phenomenon and the second mechanism was used. The numerical calculation results supporting this statement are presented in supplementary material 6.

In principle, the mechanisms involving two or more types of traps can be suggested. We don't consider such mechanisms in the present work.

\subsection{Further model developments}
The proposed model can be developed by taking into account the following factors:\\

{\bf Electric field influence}

In this work, we used model parameters such that the induced electric field is weak enough to be neglected in the calculations. In general case one should solve the joint system of equations (\ref{SRH+}-\ref{field_eq}), that accounts for electric field.
 Differences in the diffusion coefficients of holes and electrons, as well as changes in kinetic constants, can significantly distort the profile of charge carriers.  The presence of non-uniformly distributed shallow traps can also influence the electric field as well. In addition, the presence of an internal field caused by the movements of the lattice ions can also largely distort the electric field profile.
The solution of complete equations that take into account motion in the field is the subject of future study.\\

{\bf Defect mobility}

It is well known that ion migration is strongly associated with the formation of deep traps, which are associated with the change in charge recombination kinetics \cite{Panzer2017, McGovern2020, Lai2018, Garcia2022, Oranskaia2018, Chen2016, konstantinos_papadopoulos__2024,Che2019}. Moreover, this dynamic is expected to depend on the electric field \cite{Chen2020ADV}.
Thus, it may be necessary to take trap mobility into account in the model. It can be done by introducing the diffusion term into the Eq.(\ref{DefDyn}).\\

{\bf Photon recycling}

 The photon recycling effect may be significant for MAPbBr\textsubscript{3} samples \cite{deQuilettes2022, YAMADA2020116987, Behera2022, Fang2017}.
We do not take this effect into account in present work, since the PL quantum yield is considered to be very small.
Nevertheless, in the case of a large quantum yield, this process may also be added to the model.\\

{\bf Pulsed excitation}

A stationary system of equations (\ref{SRHsimple}) does not describe PL decay kinetics measured in the pulsed excitation experiment.
In order to reproduce the results of experiments on time-resolved photon counting, it is necessary to solve a non-stationary system (\ref{SRH+}),
as it was done in  Ref. \onlinecite{Kiligaridis2021} or Ref. \onlinecite{Maiberg2014}.

Because the deep traps density is an adiabatic parameter for the kinetics of charge carriers, there must be an intermediate time scale that is much slower than the recombination times of charge carriers, but at the same time much faster than the characteristic times of defect dynamics. In this case, the equations for the dynamics of the defects should include the time-averaged carrier density.

In the original work, in addition to the long-range exponential tail, a very fast nonexponential process was also observed in the decay of luminescence.  Most likely, this effect may appear due to fast trapping in shallow traps \cite{Musiienko2020, Wang2017, Yuan2024}, exciton formation and annihilation \cite{Baranowski2020}, or Auger recombination.
Shallow traps are not taken into account in our model because they practically do not affect the stationary solution for the distribution of charge carriers. If shallow traps are associated with surface defects their number and dynamics can depend on the atmosphere. In the original experiments on photoswitching, a constant quenching of luminescence was detected depending on the atmosphere. However, this quenching does not affect the fast photoswitching process. Furthermore, the dynamics of shallow traps can lead to special effects in single-photon counting statistics for sub-micrometer perovskite crystals \cite{Eremchev2023}. Thus, in order to correctly solve the problem of luminescence time-resolved decay, it is necessary to explicitly take into account the presence of shallow traps.

\subsection{Suggested experiments}
To improve our understanding of the photoswitching mechanism and further develop the model, it is necessary to carry out the following experiments.\\

{\bf A mapping of the luminescence quantum yield depending on the pumping conditions}

This technique was recently proposed in Refs. \onlinecite{Kiligaridis2021, Rao2023}. Experiments of this type make it possible to extract charge kinetics parameters and evaluate the defect density. An interesting variant of such a technique would be to carry out the experiments in conditions of luminescence and quenched luminescence, with the help of a control light. Knowing the parameters obtained from these experiments, it would be possible to discard the assumptions we made for the charge kinetics and consider a much more accurate recombination model. Changes in photoluminescence are also known to occur when exposed to particle beams \cite{Palei2020}, as well as X-rays \cite{Armaroli2021} and UV irradiation \cite{Chu2018}. Perhaps these processes could also affect photoswitching, which is a further avenue of research.\\

{\bf Confocal excitation}

In the original work, wide-field techniques were used to excite and suppress the luminescence, which made it possible to reduce the equations to one-dimensional ones. Of particular interest is the verification of the effect in a confocal scheme, in which case the solution for the diffusion equation is very different from a one-dimensional one. Such experiments would make it possible to track the interaction mechanism between defects and charge carriers much better.

The use of an electric field in both types of experiments, as well as the measurement of the photocurrent, is also of particular interest, since such experiments also contain information necessary for a more accurate construction and verification of the model.

\section{Conclusion}

In conclusion, a theoretical model has been proposed that successfully explains the key properties of the recently discovered all-optical photoswitching of individual microcrystals of perovskite with an excess of lead.
The model includes both the kinetics of charge carrier recombination and the long-term dynamics of defects associated with deep traps within the conduction band.

\section*{Supplementary material}

Analytical solutions for some limiting cases of the general model; details of numerical simulation.

\begin{acknowledgments}

Eduard Podshivaylov acknowledges support from the Foundation for the Advancement of Theoretical Physics and Mathematics "BASIS" (22-1-5-36-1). Authors acknowledge the core funding from the Russian Federal Ministry of Science and Higher Education (FWGF-2021-0002).
\end{acknowledgments}

\section*{Author Declaration Section}

\section*{Conflict of interest}

The authors have no conflicts to disclose.

\section*{Author contributions}
Eduard Podshivaylov: writing – original draft (equal), software (lead), visualization (lead), conceptualization (equal), methodology (equal). Pavel Frantsuzov: writing – original draft (equal), conceptualization (equal),  methodology (equal), supervision (lead).

\section*{Data Availability Statement}
The data that support the findings of this study are available from the corresponding author upon reasonable request.

\bibliographystyle{unsrt}
\bibliography{References}

\end{document}